\documentclass[11pt]{article}
\usepackage[a4paper,margin=1in]{geometry}
\usepackage[T1]{fontenc}
\usepackage[utf8]{inputenc}
\usepackage{lmodern}
\usepackage{microtype}
\usepackage{booktabs}
\usepackage{graphicx}
\usepackage{amsmath}
\usepackage{array}
\usepackage{multirow}
\usepackage{pdflscape}
\usepackage{caption}
\usepackage{enumitem}
\usepackage[numbers,sort]{natbib}
\usepackage{xcolor}
\usepackage[colorlinks=true,linkcolor=blue!50!black,citecolor=blue!50!black,urlcolor=blue!50!black,backref=page]{hyperref}
\renewcommand*{\backrefalt}[4]{\ifcase #1 \or Cited on page~#2.\else Cited on pages~#2.\fi}

\newcommand*{\doi}[1]{doi: \href{https://doi.org/#1}{\nolinkurl{#1}}}
\usepackage[section]{placeins}

\setlist{nosep,leftmargin=1.5em}
\newcommand{\ndcg}{nDCG@10}

\title{Decision Models as Rerankers:\\Matching Commercial Rerank APIs at a Fraction of the Cost}
\author{Jakob Gruber\\Swietelsky AG, Linz, Austria}
\renewcommand{\today}{\number\day\space\ifcase\month\or January\or February\or March\or April\or May\or June\or July\or August\or September\or October\or November\or December\fi\space\number\year}
\date{\today}

\begin{document}
\maketitle

\begin{abstract}
Chatbots and AI agents often answer from the few passages that a search puts on top, so the order of the search results decides what they can use. Rerankers set this order. Decision models are AI models built for one type of task: given a question about some input, they answer yes or no, pick one of several options or give a score, with a probability for each answer. Because they write no text, they are fast and cheap, typically \$0.01 to \$0.10 per million input tokens. This makes them suitable for reranking. This paper tests two variants, one yes/no question per passage or one multiple-choice question per query with the candidates as options. It compares 22 such setups, built on 19 decision models from 12 vendors, with nine dedicated rerankers on 35 test sets from four benchmarks. The best decision models, Perplexity Decider v1.2 and v1.3, score as well as the best commercial rerankers tested, at 20 to 45\% of their cost. Averaged over six families of test sets, Perplexity v1.2 with one question per passage reaches an \ndcg{} of 0.540 (ranking quality, 0 to 1), as do Voyage Rerank 3 and Cohere Rerank 4 Pro. Decision models lead on reasoning-heavy queries, where 17 of the 22 setups score above the best dedicated reranker. On BEIR, dedicated rerankers lead by 0.02. Three findings matter in practice. First, with multiple choice, some models score lower when the candidates arrive shuffled, so the candidates should keep the first-stage order. Second, a high position on a public leaderboard gives no assurance. \emph{With multiple choice, the top decision model of the Artificial Analysis Decision Index on 13 November 2026 is the weakest reranker} of 14 indexed models, but third per passage. Third, some providers refuse or block queries. Google's content filter blocks 6.8\% of all queries to Gemini Decision 1, and GPT-6 Luna refuses 2.4\%.
\end{abstract}

\section{Introduction}
\label{sec:intro}

Chatbots and AI agents often answer from documents that a search finds for them, which is known as retrieval-augmented generation (RAG)~\cite{lewis2020rag}. The language model reads only the few passages that the search puts on top. If a relevant passage is not among them, the answer cannot use it. Sending more passages instead costs tokens and time. Language models also use information in the middle of a long input less well than information at its start or end~\cite{liu2024lost}. The order of the search results therefore decides much of what the model can answer. Setting this order is the job of a reranker, and it runs for every single query, so its cost matters as much as its quality.

Many search pipelines, including those for RAG, work in two stages~\cite{lin2021pretrained}. A fast first stage, such as the lexical search BM25~\cite{robertson2009bm25}, an embedding-based semantic search or a hybrid of both, returns tens or hundreds of candidates. A reranker then compares the search query with all candidates and puts the best ones on top. Many rerankers are cross-encoders, which read the query together with one candidate at a time~\cite{nogueira2019passage,nogueira2020monot5}. They are offered as open models~\cite{bgererankerv2m3,qwen3reranker8b,nvidia2026nemotron} or as commercial APIs~\cite{cohere2026rerank,voyage2026rerank,google2026ranking}. Large language models can also be used for reranking~\cite{sun2023rankgpt,pradeep2023rankzephyr}, but most of them return their judgement as generated text, which is slow and expensive~\cite{sun2023rankgpt}. The search pipeline then has to parse this text, and language models do not always follow the required output format~\cite{sun2023rankgpt}. Even a failure rate of 0.1\% means tens of failed queries a day for a search pipeline that serves tens of thousands of queries.

In September 2026, TypeSafe AI introduced a new class of models that write no text at all, which it calls \emph{System One models}~\cite{typesafe2026systemone}. Models of this class are also called \emph{decision models}, the name used in TypeSafe's documentation and by OpenRouter, which serves such models from many vendors through one request format~\cite{typesafe2026primer,openrouter2026jevguide,openrouter2026decisions}. A decision model is built for one type of task. Given a question about some input, it answers yes or no, picks one of several options or gives a score on a scale, with a probability for each answer. The first one, TypeSafe's Jev, was announced on 15 September 2026, and Jev 1.13 became the first decision model on OpenRouter three days later~\cite{typesafe2026systemone,openrouter2026jevguide,openrouter2026models}. About seven weeks after that, on 5 November, OpenRouter listed at least 29 decision models. Only eight days later, on 13 November, there were already 35, from vendors such as OpenAI, Google, Microsoft, Perplexity and Nace~\cite{openrouter2026models,perplexity2026decider,nace2026drexmax,google2026decisions}.\footnote{Counted on OpenRouter's listing of decision models, each model once, without ``latest'' aliases or free copies of paid models.} The decision models I tested have list prices of \$0.01 to \$0.10 per million input tokens, and eight of them have open weights, six of these under a permissive licence (Apache 2.0 or MIT).

Judging relevance fits the question types of decision models directly. ``Is this passage relevant to this query?'' is a yes/no question, and ``Which of these passages is the most relevant?'' is a multiple-choice question. The probability that comes back serves as the relevance score. No text has to be parsed and no model has to be trained. On OpenRouter, one request format works for most decision models, so switching to another model is mostly a matter of changing its name. Some providers need small adjustments (Section~\ref{sec:protocol}). Some tools already use decision models in this way. A preview of LiteLLM's JevRank, which builds on the pilot of this work, serves the decision models of Perplexity and TypeSafe behind LiteLLM's rerank interface~\cite{litellm2026jevrank}. Open WebUI can use a decision model to rank the tools for its tool search~\cite{openwebui2026toolsearch,openwebui2026toolsearchcommit}. Blog posts, code repositories and preprints have also compared decision models with dedicated rerankers since the first weeks after Jev's release, but none of them tested more than four decision models through an API (Section~\ref{sec:related}).

I compare 22 decision-model \emph{setups} against nine dedicated rerankers (seven hosted ones as well as two small ones that I ran on my own hardware). A setup combines one of 19 decision models with one of the two variants, A (one yes/no question per passage) or C (one multiple-choice question per query), so only decision models have setups. Three of the 19 ran in both variants, which gives 22 setups. Google's semantic ranker, a tenth dedicated reranker, was tested on only 14 of the test sets, so it is not among the nine. The 35 test sets come from four public benchmarks and contain 3,315 distinct queries. They fall into six families. These are TREC Deep Learning, BEIR, the reasoning-heavy BRIGHT, full documents, long German documents and chunks of 2,000 tokens as used in RAG. Every system reranks the same candidates. Each difference between two systems is tested with a paired bootstrap. Every request is priced at the model's public list price where one exists. A single pass of all 31 systems, the 22 setups plus the nine dedicated rerankers, over all 35 test sets costs \$489 at list prices. Of this, \$60 ran without a fee, on free tiers, a free starting credit, a free launch offer or my own hardware.

The main findings are:

\begin{enumerate}
\item \textbf{The best decision models are as good as the best commercial rerankers tested, at a fraction of the cost.} Averaged over the six families (the \emph{family mean}), Perplexity Decider v1.2, asked one yes/no question per passage, reaches an \ndcg{} of 0.540 (the standard measure of ranking quality, on a scale from 0 to 1), the same as Voyage Rerank 3 and Cohere Rerank 4 Pro to three decimals. Its successor Perplexity Decider v1.3, asked one multiple-choice question per query, reaches 0.538, which is level with them, and it has the highest average over all 35 test sets. Averaged over the 35 sets, reranking 1,000 queries costs \$1.27 with Perplexity v1.2 and \$0.99 with v1.3. The same queries cost \$2.84 with Voyage Rerank 3 and \$5.05 with Cohere Rerank 4 Pro (Section~\ref{sec:results}).
\item \textbf{Each kind of system has its strength.} Decision models lead on the reasoning-heavy BRIGHT benchmark, where 17 of the 22 setups score above the best dedicated reranker. Dedicated rerankers lead on BEIR by about 0.02, as well as on long German documents read up to 8,192 tokens, by a similar margin. Microsoft Decision 1, a decision model, leads on TREC DL. On chunks of 2,000 tokens the best systems of both kinds are level. On full documents, Perplexity v1.2 with one question per passage is ahead of the best dedicated reranker by a small margin (Section~\ref{sec:results}).
\item \textbf{On long German documents, a reranker has to read far into each document.} With only the first 1,024 tokens of documents of about 8,400 tokens, every system ranked them worse than a good first stage, whether BM25 or an embedding model. With the first 8,192 tokens, 23 of the 31 systems did significantly better than the first stage with the embedding model (Section~\ref{sec:results}). Google's semantic ranker, which reads 1,024 tokens per document, beat none of these good first stages (Section~\ref{sec:google}).
\item \textbf{The prompt matters little, at least for Perplexity v1.2 and Perplexity Flash v1.2.} Five other prompts for the multiple-choice question change their result by about 0.01 or less (Section~\ref{sec:prompt}).
\item \textbf{With multiple choice, some models lean on the order in which the candidates arrive.} With the candidates of SciFact shuffled, Perplexity v1.2 loses 0.130, its successor v1.3 only 0.049. Drex Max v1.1, a decision model from Nace, loses \textbf{two thirds} of its result there (0.525 to 0.180). Gemini Decision 1, in contrast, barely changes (Section~\ref{sec:position}). In a search pipeline, leaning on the order does no harm, because the first stage already sorts the candidates. A decision model can therefore be added behind it without changes.
\item \textbf{A high position on a public leaderboard gives no assurance.} Drex Max v1.1, the top decision model of the Artificial Analysis Decision Index on 13 November 2026, was the weakest reranker of the 14 indexed models tested when asked with multiple choice, the cheaper variant that needs one request per query. Asked one yes/no question per passage instead, it ranked third of the 14, but cost about 27\% more than with multiple choice (Section~\ref{sec:aa}).
\item \textbf{Some providers refuse or block queries.} GPT-6 Luna refuses 2.4\% of the queries, while Google's content filter blocks 6.8\% of all queries to Gemini Decision 1 because of words in the candidates, although these come from public, harmless and well-known benchmarks. Both models were asked with multiple choice, so all candidates of a query went into one request. A refusal or block therefore loses the whole query, which scores 0 in all main results (Section~\ref{sec:refusals}).
\item \textbf{The field of decision models moves very fast.} Perplexity released new versions of both its models while the experiments were running. Perplexity Decider Flash v1.3 gained 0.022 over Perplexity Flash v1.2, while Perplexity v1.3 changed only by a small margin (Section~\ref{sec:results}).
\end{enumerate}

\section{Background and related work}
\label{sec:related}

\paragraph{Rerankers.} A cross-encoder reads the query together with one candidate, then outputs a relevance score. For a list of candidates, it repeats this for each one. With pretrained transformers, this line of models starts with BERT~\cite{nogueira2019passage}, followed by sequence-to-sequence models~\cite{nogueira2020monot5}. It continues with small models such as MiniLM-L6~\cite{wang2020minilm,minilmcrossencoder}, multilingual models such as bge-reranker-v2-m3~\cite{bgererankerv2m3} (which starts from the M3 encoder~\cite{chen2024bgem3}) and rerankers built on language models with several billion parameters~\cite{zhang2025qwen3embedding,qwen3reranker8b,nvidia2026nemotron}. Commercial rerankers take a query with a list of candidates and return the candidates sorted by relevance~\cite{cohere2026rerank,voyage2026rerank,google2026ranking}. The model stays behind the API, so its size is not public. All of these models are trained for relevance.

\paragraph{Language models as rerankers.} Large language models can also rerank, either zero-shot or after fine-tuning for ranking. Pointwise methods score one candidate at a time, for example by the likelihood that the model answers yes to a relevance question~\cite{liang2023helm}, or by the expected relevance over several graded labels~\cite{zhuang2024beyond}. Listwise methods show the model several candidates and let it write out a ranking, with a sliding window for long lists~\cite{sun2023rankgpt,ma2023listwise}. RankVicuna and RankZephyr are open models with 7B parameters, fine-tuned on MS MARCO queries to reproduce rankings written by GPT-3.5. RankZephyr then also learned from rankings written by GPT-4~\cite{pradeep2023rankvicuna,pradeep2023rankzephyr}. FIRST is a listwise model fine-tuned for ranking. In each window, the candidates carry labels, and FIRST ranks them by the score the model gives each label as the first token of its answer~\cite{reddy2024first}. Pairwise methods compare two candidates at a time~\cite{qin2024prp}. Setwise methods ask for the most relevant candidate of a small set~\cite{zhuang2024setwise}. For reasoning-heavy queries, rerankers such as Rank1 first write out their reasoning~\cite{weller2025rank1}. All of these methods trade quality against the number of model calls and the amount of text the model writes. Compared with cross-encoders, rerankers built on large language models can be slower and more costly~\cite{dejean2024thorough}. Through an API they are also costly, for example \$0.596 per query for RankGPT with GPT-4~\cite{sun2023rankgpt}, and their text output is not always well formed~\cite{sun2023rankgpt}.

The two variants in this paper map onto this line of work. The yes/no question is pointwise, since the probability of yes is the score of a candidate. The multiple-choice question is closest to the setwise idea, applied once to the whole candidate list. The probability of each option is the score of that candidate. One listwise method of the Setwise paper~\cite{zhuang2024setwise} scores candidates the same way, with the probability that the model writes the label of each candidate. What is new is where this probability comes from. A general decision model returns this probability as its normal output for any question, whether through an API or on local hardware, so no access to the token probabilities of the model is needed. Several of the methods above read the internal probabilities that a model assigns to certain words, for example to ``yes''~\cite{liang2023helm,zhuang2024beyond,qin2024prp,zhuang2024setwise,reddy2024first}. Pairwise and setwise prompting can also work from the generated answer, which then has to be parsed~\cite{qin2024prp,zhuang2024setwise}. With the decision models tested here, nothing is generated or parsed, and every one of them with an API price charges only for reading the input.

\paragraph{Position effects.} Language models do not treat all parts of a long input alike. For example, they often use information at the start or the end of a long input better than information in its middle~\cite{liu2024lost}. They prefer certain option labels in multiple-choice questions~\cite{zheng2024selection}. Their listwise rankings depend on the input order~\cite{tang2024found}. A multiple-choice question with 100 options invites all three effects. Section~\ref{sec:position} tests how much the candidate order changes the result.

\paragraph{Benchmarks.} TREC Deep Learning~\cite{craswell2020dl19,craswell2021dl20} and BEIR~\cite{thakur2021beir} are standard retrieval benchmarks~\cite{bright2509}. As in RankGPT~\cite{sun2023rankgpt}, this paper reranks the BM25 top 100. BRIGHT~\cite{su2025bright} adds queries whose relevant documents are hard to find by keyword or semantic matching alone, with reproducible baselines in~\cite{bright2509}. MLDR~\cite{chen2024bgem3} provides long documents in 13 languages, of which this paper uses German.

\paragraph{Decision models.} OpenRouter serves decision models from many vendors through one request format~\cite{openrouter2026decisions,openrouter2026jevguide}. Several decision models are documented in model cards and papers~\cite{perplexity2026decider,perplexity2026flash,startlux2026,nace2026drexmax}. Artificial Analysis, an independent benchmarking company, publishes the Artificial Analysis Decision Index, a public leaderboard that ranks decision models by general decision benchmarks~\cite{aa2026decisionindex}. Two tools already use decision models for retrieval-like tasks. A preview of LiteLLM serves the models of two vendors through its rerank interface~\cite{litellm2026jevrank}, and Open WebUI can use one to pick which tools to hand to the agent (tool search)~\cite{openwebui2026toolsearch}. Both rely on the assumption that a decision model judges relevance well, which LiteLLM checked for Jev on SciFact.

\paragraph{Decision models as rerankers.} Within weeks of Jev's release, blog posts and code repositories compared Jev with cross-encoders or commercial rerankers on sets from BEIR, TREC DL or BRIGHT. There, its average came within about 0.01 of the best of them or beat it~\cite{belbati2026jev,hevenor2026jev,huang2026jevqwen,villarini2026s1rank,vvrrao2026jevbench,hevenor2026company}. On multilingual queries from MIRACL, it trailed the best by about 0.06, while on Japanese queries from JQaRA it beat the best of the four rerankers run there by about 0.07~\cite{vkehfdl2026jevreranker}. A preprint then found that Jev as well as Perplexity Decider v1.1 rank ahead of three open cross-encoders on average over eight BEIR sets~\cite{zhuang2026beyondranking}. Another found that up to 9B parameters, a dedicated reranker ranks 0.05 to 0.29 higher in \ndcg{} than a decision model built on the same open base model, while at 27B the two tie~\cite{dinzinger2026notbetter}. Two more ran decision models against dedicated rerankers on TREC DL, BEIR or BRIGHT~\cite{wu2026decisionrank,li2026replace}. In one of them~\cite{li2026replace}, Perplexity v1.1 on local hardware failed on more than half of the BRIGHT queries unless each candidate was shortened. Each of these studies tested at most four decision models through an API. Two of the preprints ran 10 and 12 decision models, all of them open except Jev in the first~\cite{dinzinger2026notbetter,li2026replace}. This paper tests 18 through an API as well as one on my own hardware, all on the same 35 test sets. I began this work in September 2026, independently of these studies.

\section{Using a decision model as a reranker}
\label{sec:method}

A decision model takes two inputs. The first is a \emph{state}, usually any JSON object, and the second is one or more \emph{questions} about that state. For each question it returns an answer with a probability. The common question types are yes/no, multiple choice and a score on a scale. Some providers add others, such as image bounding boxes or text spans~\cite{openrouter2026decisions,perplexity2026changelog}. This paper uses the first two. A yes/no question returns the probability of yes, a number between 0 and 1. A multiple-choice question lists named options and returns one probability per option.

Reranking fits this directly. A first stage returns candidates $d_1,\dots,d_n$ for a query $q$. The reranker needs one relevance score $s_i$ per candidate, then sorts the candidates by it. The paper tests two variants of the request, shown in Figure~\ref{fig:variants}. In the tables, variant A is the yes/no question per passage and variant C the multiple-choice question per query. Variants B and D were dropped after a pilot, as the end of this section explains.

\paragraph{Variant A (one yes/no question per passage).} The state holds the query and one candidate. The question is \emph{``Is the passage relevant to the query, meaning it contains information that answers, supports or contradicts it?''} The probability of yes is the score $s_i$. A reranking task with $n$ candidates therefore sends $n$ requests to the decision model. They are independent of each other, so they can be sent in parallel. A score depends neither on the other candidates nor on their order.

\paragraph{Variant C (one multiple-choice question per query).} The state holds the query and all $n$ candidates, labelled \texttt{p0} to \texttt{p}$_{n-1}$ in first-stage order. The question is \emph{``Which passage is the most relevant to the query?''}, with one option per label. The probability of option \texttt{p}$_{i-1}$ is the score $s_i$. A query needs only one request, so the query text is paid for once. If the provider rejects the request, in most cases because it is too large, that query falls back to one question per passage.

\begin{figure}[!htb]
\small
\begin{minipage}[t]{0.47\textwidth}
\textbf{Variant A} (one request per passage)
\begin{verbatim}
model: <decision model>
state:
  query: <query>
  passage: <one passage>
questions:
  rel:
    type: noul        # yes/no
    instructions: >
      Is the passage relevant to
      the query, meaning ...?
\end{verbatim}
\end{minipage}\hfill
\begin{minipage}[t]{0.49\textwidth}
\textbf{Variant C} (one request per query)
\begin{verbatim}
model: <decision model>
state:
  query: <query>
  passages:
    p0: <first passage>
    p1: <second passage>
    ...
questions:
  best:
    type: choice
    instructions: >
      Which passage is the most
      relevant to the query?
    criteria:
      p0: null
      p1: null
      ...
\end{verbatim}
\end{minipage}
\caption{The two request formats, written as YAML for readability. The API receives the same structure as JSON. In variant A, the probability of yes is the score of the candidate. In variant C, the probability of each option is the score of that candidate. \texttt{noul} is the API's name for a yes/no question. The \texttt{criteria} list the options, and \texttt{null} means an option has no description beyond its label.}
\label{fig:variants}
\end{figure}

In both variants the candidates are then sorted by $s_i$, highest first. Ties keep the first-stage order. Ties are common in practice. Some providers round their probabilities. In variant C, the probabilities of all options add up to 1. With many options, a few take most of the probability, so some models give many candidates the same score close to zero. Section~\ref{sec:position} measures how much the tie rule adds with multiple choice.

\paragraph{Two dropped variants.} A pilot also tested two further variants, B as well as D, with the top 20 candidates of a separate BM25 run on SciFact (300 queries, passages not cut) and on 100 queries with long German documents. Like variant C, both send one request per query, but like variant A, they ask a yes/no question for each candidate (Figure~\ref{fig:variants-bd}). In variant B, each question carries the text of its candidate. In variant D, the state holds all candidates, labelled as in variant C, and each question names one label. In these tests, variant B scored at most level with variant C and was slower. GPT-6 Luna refused 19 of the 300 SciFact queries with it, against none with A or C, although SciFact is a public set of scientific claims. The refusals repeated on every retry. Variant D was never better than variant C for Perplexity v1.2. For GPT-6 Luna it was ahead only on the German documents cut to 1,024 tokens. There it gave the 20 candidates of a query only about two different scores on average, so the first-stage order decided most of its ranking. In the pilot, variant D cost 4 to 45\% more than C, because its one question per candidate adds input tokens.

\begin{figure}[!htb]
\small
\begin{minipage}[t]{0.47\textwidth}
\textbf{Variant B} (one request per query)
\begin{verbatim}
model: <decision model>
state:
  query: <query>
questions:
  d0:
    type: noul        # yes/no
    instructions: >
      Is the passage relevant to
      the query, meaning ...?
      Passage: <first passage>
  d1:
    type: noul
    instructions: >
      ... Passage: <second passage>
  ...
\end{verbatim}
\end{minipage}\hfill
\begin{minipage}[t]{0.49\textwidth}
\textbf{Variant D} (one request per query)
\begin{verbatim}
model: <decision model>
state:
  query: <query>
  passages:
    p0: <first passage>
    p1: <second passage>
    ...
questions:
  d0:
    type: noul        # yes/no
    instructions: >
      Is passage p0 relevant to
      the query, meaning ...?
  d1:
    type: noul
    instructions: >
      Is passage p1 relevant ...?
  ...
\end{verbatim}
\end{minipage}
\caption{The two request formats that the pilot dropped, written as YAML like Figure~\ref{fig:variants}. Both send one request per query with one yes/no question per candidate. Each question returns the probability of yes, which is the score of its candidate.}
\label{fig:variants-bd}
\end{figure}

Both prompts were fixed before the experiments and never tuned. Section~\ref{sec:prompt} tests five other prompts for the multiple-choice question. They change the result by about 0.01 \ndcg{} or less.

\paragraph{What the variants cost.} Every tested decision model with an API price is priced per input token. Variant A repeats the query and the question with every candidate. Variant C sends them only once. On TREC DL, DBPedia and Signal-1M, whose passages are the shortest, this makes variant A about twice as expensive or more, while on the other six of the eight BEIR sets it costs 1.2 to 1.3 times as much. On BRIGHT, whose queries are often long, it costs 1.4 to 3.2 times as much, because the query is repeated with every candidate. On TREC DL 2019, with 100 passages of at most 512 tokens per query, Perplexity v1.2 reads 20,138 input tokens per query in variant A and 9,966 in variant C. That is \$0.40 against \$0.20 per 1,000 queries. On Signal-1M, whose passages are short posts, variant A costs about four times as much (4.1 times with Perplexity v1.2, 3.8 with v1.3). On long documents the difference nearly disappears. On the German sets cut at 8,192 tokens and on the full documents, variant A costs only 1 to 7\% more, while on the German sets cut at 1,024 tokens it still costs 10 to 18\% more. Over all German sets, Perplexity v1.2 costs \$1.63 per 1,000 queries with one question per passage and \$1.59 with multiple choice. Multiple choice is thus never the more expensive variant, and on the shortest passages it costs about half as much or less.

The variants also differ in how they run. Variant A needs $n$ requests per query, so it reaches a provider's limit on requests per minute sooner than variant C. Its latency is the time until the last of the $n$ requests is answered, which is the slowest request when all of them are sent at once. Variant C needs a single request, whose latency grows with the total input. It also needs an input limit large enough for all candidates at once.

\section{Experimental design}
\label{sec:setup}

\subsection{Test sets}

The 35 test sets fall into six families (Table~\ref{tab:bench}). A set is a list of queries, each with a fixed list of candidates from a first stage. Every system reranks exactly the same candidates. The one exception is Google's semantic ranker, which received the German documents in full length (Section~\ref{sec:google}). Together the sets hold 6,003 queries. Many of them appear in more than one set, for example once with short and once with long documents, so only 3,315 of them are different from each other.

\begin{table}[!htb]
\centering
\footnotesize
\setlength{\tabcolsep}{3.5pt}
\resizebox{\textwidth}{!}{%
\begin{tabular}{lrrlll}
\toprule
Family & Sets & Queries & Content & First stage & Cut (tokens) \\
\midrule
TREC DL & 2 & 97 & English web passages & BM25, top 100 & 512 \\
BEIR & 8 & 1,510 & English, six domains & BM25, top 100 & 512 \\
BRIGHT & 12 & 1,384 & English, reasoning-heavy & BM25, top 100 & 512 \\
Full documents & 2 & 306 & English news articles & BM25, top 100 & none \\
Long German documents & 8 & 2,000 & German, median 8,390 tokens & BM25 and two embedding models, top 20 &1,024 / 8,192 \\
Chunks & 3 & 706 & German and English & embedding model, top 20 chunks & 2,000 \\
\bottomrule
\end{tabular}}
\caption{The 35 test sets in six families. Queries are counted once per set, so the German family counts its queries once per first stage and cut. Cut: maximum tokens kept per candidate.}
\label{tab:bench}
\end{table}

\paragraph{TREC DL and BEIR.} The setup follows the protocol of RankGPT~\cite{sun2023rankgpt}, with the passage ranking tasks of TREC Deep Learning 2019 and 2020~\cite{craswell2020dl19,craswell2021dl20} on MS MARCO~\cite{bajaj2016msmarco}, plus the same eight BEIR sets~\cite{thakur2021beir}. These are TREC-COVID, NFCorpus, Touch\'e-2020, DBPedia, SciFact, Signal-1M, TREC-News and Robust04. The first stage is BM25~\cite{robertson2009bm25} as implemented in Pyserini, a standard toolkit for search experiments~\cite{lin2021pyserini}, with its default parameters ($k_1=0.9$, $b=0.4$). It runs on the ready-made Pyserini indexes and returns the top 100 documents. For BEIR, these are the flat indexes, which put the title and text of a document into one field. Every candidate is cut to its first 512 tokens, counted with the \texttt{o200k\_base} tokenizer. The BM25 results over all queries are within 0.01 of the values published for Pyserini on every set~\cite{pyserini2cr-msmarco,pyserini2cr-beir}. TREC DL 2019 has 43 judged queries and TREC DL 2020 has 54. On NFCorpus, BM25 returns no candidate for 15 of the 323 queries. They are left out for every system, which raises the BM25 result of the set from 0.322 to 0.337.

\paragraph{BRIGHT.} The BRIGHT family uses all twelve sets of BRIGHT~\cite{su2025bright}, where finding the relevant document takes reasoning beyond shared words or shared meaning. As for BEIR, BM25 returns the top 100 documents, each cut to its first 512 tokens. AoPS, LeetCode and TheoremQA-Questions exclude certain documents for each query, as the benchmark requires, so most of their queries have fewer than 100 candidates. In total, 525 of the 2,991 queries of TREC DL, BEIR and BRIGHT have fewer than 100 candidates. Of these, 415 belong to the three sets above and 110 to BEIR (108 of them to NFCorpus). The BM25 first stage reaches a mean of 0.137, the same as the plain BM25 baseline published by Sharifymoghaddam et al.~\cite{bright2509}. Here the first stage limits what any reranker can reach. On the two TheoremQA sets, the BM25 top 100 contains a relevant document for only 18 to 20\% of the queries. The two sets share their queries, since all 76 of TheoremQA-Theorems also belong to TheoremQA-Questions, with different candidates.

\paragraph{Full documents.} TREC-News and Robust04 consist of long news articles. Both sets run a second time with the same BM25 candidates but without the cut, so each system reads as much of every document as its input limit allows (Section~\ref{sec:protocol}).

\paragraph{Long German documents.} The German part of MLDR~\cite{chen2024bgem3} has 10,000 documents with a median length of 8,390 tokens. Of these, 78\% are longer than 1,024 tokens, while 52\% are longer than 8,192. The sets use the 400 queries of the test and development splits. Their queries were generated from paragraphs anywhere in a document~\cite{chen2024bgem3}, so the answer can lie far from the start. The 1,847 training queries are left out, because their labels do not fit the corpus. For 992 of them, the labelled document ID does not occur in it. For the other 855, the ID does occur, but it belongs to a document with a different text, so the label points to the wrong document. Four first stages each return the top 20 documents. BM25 with German stemming and stop words runs on two disjoint samples of 100 queries (recall@20 of 0.700 for the first, 0.730 for the second). Two embedding models run on all 400 queries. The first, mistral-embed, has an input limit of 8,192 tokens of its own tokenizer and had to cut 7,131 of the 10,000 documents to it (recall@20 of 0.368). The second, Qwen3 Embedding 8B~\cite{zhang2025qwen3embedding}, was run with documents cut at 30,000 tokens, below its input limit of 32K tokens, which affected only 239 documents (recall@20 of 0.705). Every candidate list is reranked twice, once with the documents cut to their first 1,024 tokens and once to their first 8,192. These eight sets show how much of a long document a reranker has to read, and whether reranking still helps when the first stage is already good.

\paragraph{Chunks.} This family models the input of a RAG pipeline. The documents of MLDR German, TREC-News and Robust04 are split into chunks of 2,000 tokens. For each query, Qwen3 Embedding 8B ranks the chunks of its candidate documents, which are the 20 documents of the Qwen3 Embedding 8B first stage for MLDR and the top 30 BM25 documents for the two English sets. The top 20 chunks go to the rerankers. A chunk inherits the relevance label of its document, because the benchmarks label whole documents.

\subsection{Systems}
\label{sec:systems}

\paragraph{Decision models.} Table~\ref{tab:systems} lists all systems. The decision models come from the OpenRouter listing of 5 November 2026~\cite{openrouter2026models}. Twelve of them went through a pilot on SciFact, which reranked the top 20 candidates of a separate BM25 run with uncut passages. A model entered the main runs only if it improved on BM25 alone (0.686 \ndcg{}) by at least 0.05. Nine models did, while Upstage Solar Decide (0.651), Cloudflare Clef (0.710) and Respan Span 02 (0.719) did not. Five models that were not part of the pilot entered without screening, namely Liquid d1.1, Baseten joey-1, Baseten elephant-1, Upstage Solar Decide Flash and Morph Reflex. Models that do not accept a free-form state were left out, and seven other listed models were not run.

\begin{table}[!htb]
\centering
\footnotesize
\setlength{\tabcolsep}{3pt}
\resizebox{\textwidth}{!}{%
\begin{tabular}{llclllrr}
\toprule
Vendor & Model & Variant & Weights & Parameters & Access & \$/1k (DL19) & Latency (s) \\
\midrule
\multicolumn{8}{l}{\emph{Decision models}} \\
Perplexity & Decider v1.3 & C & open, Apache 2.0 & 26.1B & OpenRouter, free launch offer & 0.20 & 1.20 \\
Perplexity & Decider v1.3 & A & open, Apache 2.0 & 26.1B & OpenRouter, free launch offer & 0.38 & -- \\
Perplexity & Decider Flash v1.3 & C & open, Apache 2.0 & 4.54B & OpenRouter, free launch offer & 0.10 & 0.39 \\
Perplexity & Decider v1.2 & C & open, Apache 2.0 & 26.1B & OpenRouter & 0.20 & 0.61 \\
Perplexity & Decider v1.2 & A & open, Apache 2.0 & 26.1B & OpenRouter & 0.40 & -- \\
Perplexity & Decider Flash v1.2 & C & open, Apache 2.0 & 4.54B & OpenRouter & 0.10 & 0.42 \\
OpenAI & GPT-6 Luna (Decisions) & C & closed & -- & OpenRouter & 0.93 & 0.42 \\
Liquid & d1.1 & C & closed & -- & OpenRouter & 0.19 & 0.63 \\
Liquid & d1-ultra & A & closed & -- & OpenRouter & 0.63 & -- \\
Liquid & d1 & A & closed & -- & OpenRouter & 0.30 & -- \\
TypeSafe & Jev 1.14 & A & closed & -- & OpenRouter & 1.67 & -- \\
Microsoft & Decision 1 & A & closed & -- & OpenRouter & 0.57 & -- \\
Inception & Mercury Decide & A & closed & -- & OpenRouter & 0.69 & -- \\
Morph & Reflex & A & closed & -- & OpenRouter & 0.88 & -- \\
Baseten & elephant-1 & A & open, MIT & 9.41B & OpenRouter & 0.78 & -- \\
Baseten & joey-1 & A & open, MIT & 2.72B & OpenRouter & 0.22 & -- \\
Nace & Drex v2.0 & A & open, Open RAIL-M & 35.1B (MoE) & OpenRouter & 0.78 & -- \\
Nace & Drex Max v1.1 & C & closed & -- & vendor API, free starting credit & 0.30 & 0.87 \\
Nace & Drex Max v1.1 & A & closed & -- & vendor API, free starting credit & 0.60 & -- \\
Google & Gemini Decision 1 & C & closed & -- & Google AI Studio, OpenRouter & 0.96 & 0.41 \\
Upstage & Solar Decide Flash & A & closed & -- & OpenRouter & 4.85 & -- \\
StartLux & StartLux-Decision v1.1 9B Instant$^{\S}$ & A & open, CC BY-NC 4.0 & 9.65B & self-hosted & 0.00 & -- \\
\midrule
\multicolumn{8}{l}{\emph{Dedicated rerankers}} \\
Cohere & Rerank 4 Pro & -- & closed & -- & OpenRouter & 2.50 & 0.70 \\
Cohere & Rerank 4 Fast & -- & closed & -- & OpenRouter & 2.00 & 0.79 \\
Cohere & Rerank 3.5 & -- & closed & -- & OpenRouter & 1.00 & 0.67 \\
Voyage AI & Rerank 3 & -- & closed & -- & OpenRouter & 0.45 & 0.52 \\
Voyage AI & Rerank 2.5 & -- & closed & -- & OpenRouter & 0.45 & 0.42 \\
Alibaba & Qwen3 Reranker 8B & -- & open, Apache 2.0 & 8.19B & OpenRouter & 3.32 & 1.71 \\
NVIDIA & Nemotron 3.5 Rerank 8B & -- & open, OpenMDW 1.1 & 8.38B & OpenRouter, free tier & 0.50 & 1.95 \\
BAAI & bge-reranker-v2-m3 & -- & open, Apache 2.0 & 568M & self-hosted & 0.11 & -- \\
-- & MiniLM-L6 cross-encoder$^{\S}$ & -- & open, Apache 2.0 & 22.7M & self-hosted & 0.00 & -- \\
Google & semantic-ranker-default (v005) & -- & closed & -- & Vertex AI, via a proxy & 1.00 & 0.43 \\
\bottomrule
\end{tabular}}
\caption{The systems, grouped by kind and by vendor. Variant: C is one multiple-choice question per query, A is one yes/no question per passage (Section~\ref{sec:method}). Perplexity Decider v1.2 and v1.3 as well as Drex Max v1.1 ran in both. Weights: licence of the published weights, or closed when none was found in November 2026. \$/1k is the cost per 1,000 queries on TREC DL 2019 (100 passages of at most 512 tokens per query) at list price. Latency is the median time per query on the same set for systems that need one request per query, measured from one client in Europe without control of load or region. Google's semantic ranker ran on 14 of the 35 sets. $^{\S}$\,self-hosted, no known public price.}
\label{tab:systems}
\end{table}

Five more models were added later. Nace Drex Max v1.1 is a closed model that the Drex technical report mentions without saying how it was built~\cite{nace2026drexmax}. It led the decision models on the Artificial Analysis Decision Index on 13 November 2026~\cite{aa2026decisionindex} and ran through the vendor's own API. StartLux-Decision v1.1 9B Instant~\cite{startlux2026,startlux2026hf} ran on my own hardware. Google released Gemini Decision 1~\cite{google2026decisions} on 11 November 2026~\cite{google2026releasenotes}, and it ran through Google AI Studio as well as through OpenRouter. For the same request, the two routes return the same probabilities. Perplexity released Decider v1.3 together with Decider Flash v1.3~\cite{perplexity2026decider13,perplexity2026flash13,perplexity2026changelog} in November 2026, with the same sizes, licence and list prices as their predecessors~\cite{perplexity2026decider,perplexity2026flash,perplexity2026decider13,perplexity2026flash13,perplexity2026pricing}. OpenRouter listed both on 12 November 2026~\cite{openrouter2026models}, and both ran there. This gives 19 models from 12 vendors. The main grid ran from 3 to 6 November 2026. StartLux v1.1 9B followed from 8 to 10 November, Drex Max v1.1 on 9 November, Gemini Decision 1 on 11 November, Perplexity v1.3 and Perplexity Flash v1.3 on 13 November. All listings, prices and leaderboard positions in this paper are as of 13 November 2026, the last day of the main runs. Only the per-passage run of Drex Max v1.1 came later.

In the main runs, eight models ran with multiple choice (variant C) and the other eleven with one yes/no question per passage (variant A), which every decision model supports. The eight are Perplexity v1.2, Perplexity Flash v1.2, GPT-6 Luna, Perplexity v1.3, Perplexity Flash v1.3, Liquid d1.1, Drex Max v1.1 as well as Gemini Decision 1. For the first three, the pilot had compared variants A to C, plus variant D for Perplexity v1.2 and GPT-6 Luna. Multiple choice was the best variant or level with the best for Perplexity v1.2 as well as Perplexity Flash v1.2. GPT-6 Luna scored higher with it than with one question per passage on SciFact (0.773 against 0.751), but slightly lower on German at 1,024 tokens (0.516 against 0.523), where four of its multiple-choice queries were refused. For Gemini Decision 1, a check on SciFact confirmed the choice, with 0.800 for multiple choice against 0.772 for one question per passage (difference 0.028 [0.009, 0.048]). Perplexity v1.3 as well as Perplexity Flash v1.3 ran like their predecessors, while Liquid d1.1 and Drex Max v1.1 ran with multiple choice without a check of the variants. Perplexity v1.2 and v1.3 also ran with one question per passage on all 35 sets, which allows a direct comparison of the two variants. After its weak result with multiple choice, Drex Max v1.1 ran with one question per passage as well, from 16 to 18 November. This gives 22 setups in total.

\paragraph{Dedicated rerankers.} Seven rerankers ran through hosted APIs. Five of them are the commercial products Cohere Rerank 4 Pro, Cohere Rerank 4 Fast, Cohere Rerank 3.5~\cite{cohere2026rerank}, Voyage Rerank 3 and Voyage Rerank 2.5~\cite{voyage2026rerank}. The other two are the open-weight models Qwen3 Reranker 8B~\cite{qwen3reranker8b} and NVIDIA Nemotron 3.5 Rerank 8B~\cite{nvidia2026nemotron}. OpenRouter also listed Voyage Rerank 3 Lite and Rerank 2.5 Lite at \$0.02 per million tokens, two fifths of the price of Voyage Rerank 3~\cite{openrouter2026rerank}. They were not tested. Two small open rerankers ran on my own hardware, bge-reranker-v2-m3~\cite{bgererankerv2m3} and the MiniLM-L6 reranker trained on MS MARCO~\cite{wang2020minilm,minilmcrossencoder}. Google's semantic ranker~\cite{google2026ranking} was available to me only through a LiteLLM proxy. It ran on 14 of the 35 sets, so it is reported separately (Section~\ref{sec:google}).

\paragraph{Prices.} Every system with a public API price is shown at that list price, also where the runs used a free offer or my own hardware. The tables therefore compare what each system costs to use. Models that ran through OpenRouter are shown at OpenRouter's price, even where the same model has a lower price elsewhere. DeepInfra, for example, lists Qwen3 Reranker 8B at \$0.05 per million tokens, a quarter of OpenRouter's \$0.20~\cite{deepinfra2026qwen}. Solar Decide Flash as well as Mercury Decide ran at a 50\% discount on OpenRouter~\cite{openrouter2026solar,openrouter2026mercury}. They are shown at their \emph{undiscounted} prices of \$0.10 and \$0.04 per million input tokens~\cite{openrouter2026solar,inception2026decisions}. Perplexity v1.3 and Perplexity Flash v1.3 ran on a free launch offer of OpenRouter. They are shown at Perplexity's list prices of \$0.02 for Perplexity v1.3 and \$0.01 for Perplexity Flash v1.3 per million input tokens~\cite{perplexity2026pricing}. Drex Max v1.1 ran on the vendor's free starting credit and is shown at its list price of \$0.03 per million input tokens~\cite{nace2026pricing}. Part of the requests to Gemini Decision 1 went to the free tier of Google AI Studio. Nemotron 3.5 Rerank 8B ran on the free tier of OpenRouter, which allows 20 requests per minute and, for an account that has bought at least 10 credits, 1,000 requests per day. It works only with the account setting that lets providers of free models train on the submitted data. OpenRouter offers no paid endpoint for it, so it is shown at \$0.05 per million tokens, the list price of the same model at DeepInfra~\cite{deepinfra2026nemotron}, with the token counts that OpenRouter reported. StartLux v1.1 9B ran on one consumer graphics card, an AMD Radeon RX 7900 XTX with 24\,GB. The bge-reranker-v2-m3 and MiniLM-L6 models ran on the same card or, for part of the sets, on the CPU. On 13 November 2026, Novita AI listed the bge-reranker-v2-m3 model as an API at \$0.01 per million input tokens~\cite{novita2026bge}, so its runs are priced at that rate, with the tokens of every query-candidate pair counted by its own tokenizer. StartLux v1.1 9B ran as the vendor's 8-bit GGUF build under llama.cpp, with the server settings its model card asks for~\cite{startlux2026hf} and one yes/no question per passage. StartLux v1.1 9B and MiniLM-L6 have no public API price that I could find. All tables mark them with $^{\S}$ and show them at \$0, which leaves the cost of hardware or power unpriced.

\subsection{Protocol}
\label{sec:protocol}

In every setup and on every set, the decision model gets the question of its variant exactly as given in Section~\ref{sec:method}. The question stays in English for the German documents as well. Hosted rerankers receive the query with the candidate texts, but not the definition of relevance that the yes/no question gives. Voyage Rerank 2.5 accepts instructions added to the query~\cite{voyage2026rerank}. They were not used. A multiple-choice request that the provider rejects, in most cases for its size, falls back to one question per passage for that query. This happened for 60 of the 6,003 queries with each Perplexity model, for 59 with Gemini Decision 1 and for 31 with Drex Max v1.1, but never with GPT-6 Luna. These queries are still answered, one passage at a time, so the fallback changes the means very little. On OpenRouter, the Perplexity models as well as Gemini Decision 1 read up to 262,144 tokens per request, GPT-6 Luna up to 1,050,000~\cite{openrouter2026models}. Drex Max v1.1 accepted requests of about 160,000 tokens on the German sets at 8,192 tokens. Yet at least 17 of its 31 fallbacks came on BRIGHT and the German chunks, whose requests were far shorter. It therefore also rejected requests for reasons other than size. Liquid d1.1, which reads up to 65,536 tokens, fell back for 1,209 queries (20\%). These include every query of the two full-document sets. On the German sets at 8,192 tokens, the fallbacks cover every query of three sets and 210 of the 400 queries of the fourth. On these six sets, the row of Liquid d1.1 therefore mostly shows one question per passage. Google's API sorts the option labels alphabetically before the model sees them, which would put \texttt{p10} before \texttt{p2}. For Gemini Decision 1, the labels were therefore zero-padded (\texttt{p00} to \texttt{p99}), so their alphabetical order is the first-stage order. Google's API and Liquid d1.1 both reject a multiple-choice question with a single option, so the 13 NFCorpus queries with a single candidate were sent to both models as one yes/no question.

On the full-document sets, a system whose input limit cannot hold a document reads the longest beginning of it that fits. When a request did not fit, the query was sent again with the documents cut to a shorter length. For Qwen3 Reranker 8B and the Voyage models, a few queries still failed (Section~\ref{sec:refusals}). Unlike a rejected multiple-choice request, a rerank request above Voyage's token limit per request got no fallback to smaller requests. Each Voyage model lost three queries this way, which could have added at most 0.0005 to its family mean. Refusals and blocked requests were final. GPT-6 Luna, for example, refused again on each of four retries. If the request for a single candidate still failed, that candidate is ranked below all scored candidates of its query.

A query that the model refuses, that the provider blocks or that still fails after the retries counts as a task not done and \textbf{scores 0}. This follows \texttt{trec\_eval} with its \texttt{-c} option, which scores a query without a ranking as 0. Two alternatives were considered and rejected. Dropping such queries would grade each system on its own subset of the queries. A system that answers everything would be measured on all queries, one that refuses some only on the rest, so the two results would no longer measure the same thing. Dropping refused queries would also reward a model for refusing. A model that refused exactly its hardest queries would raise its average, because the queries it would have ranked worst no longer count. Refusals also cluster on certain topics and sets (Section~\ref{sec:refusals}), so a result over the answered queries alone is biased in a direction nobody can measure. Keeping the first-stage order for such queries would credit the system for a ranking the model never produced. It also assumes that a prior order exists, which is not the case in uses such as tool search or the selection of memories (Section~\ref{sec:discussion}). The fair comparison is therefore between results that count every query, with the refused ones at 0. Table~\ref{tab:overview} reports the share of such queries for each system, while Table~\ref{tab:refusals} lists the refusals as well as the blocks on the sets most affected. Section~\ref{sec:refusals} gives the result that Gemini Decision 1 would reach if the first-stage order were kept for such queries.

\subsection{Scoring and statistics}

The measure is \ndcg{}~\cite{jarvelin2002ndcg}, the normalised discounted cumulative gain of the ten highest-ranked candidates, as computed by \texttt{trec\_eval}. The gain is the relevance grade, and the discount is $\log_2(\text{rank}+1)$ at every rank. The original definition with base 2 divides by $\log_2(\text{rank})$ from rank 2 on, so the two differ at every rank after the first. The result of a family is the mean of its sets. The paper reports two overall means, because they do not rank the systems in the same order. The \emph{family mean} gives each of the six families one sixth of the weight. The \emph{set mean} is the plain mean of the 35 sets, in which BRIGHT alone accounts for 12 sets but TREC DL for only two.

To check whether a difference between two systems could be due to chance, the paper uses a paired bootstrap~\cite{sakai2006bootstrap,smucker2007significance}. It draws the queries of each set 5,000 times at random with replacement, computes the difference of the two systems on each draw, and takes the middle 95\% of these differences as the interval~\cite{efron1993bootstrap}. If the interval excludes zero, the difference is unlikely to be due to chance. Two systems are called level when their difference is 0.004 \ndcg{} or less, and one is ahead by a small margin when the difference is 0.005 to 0.009. A difference of 0.01 or more is significant when its interval stays on one side of zero. Otherwise, the leader is ahead, but not significantly. The tables give the interval of every difference.

\subsection{Cost and latency}

The cost of a request is what the provider reported, without platform fees and tax, or the cost at list price where a run was free (see above). Cost is given in US dollars per 1,000 queries, except for the totals of one pass over the 35 sets. The cost of a family is the mean over its sets, while the overall cost is the mean over the 35 sets. Latency was recorded for every request from a single client in Europe, so it is reported only as an indication (Section~\ref{sec:latency}). For systems with one request per passage, the latency of a query is that of its slowest request.

\section{Results}
\label{sec:results}

\subsection{Overall}

Table~\ref{tab:overview} shows all 31 systems on all 35 sets. Reranking helps a lot. The two best systems, Voyage Rerank 3 and Perplexity v1.2 per passage, each raise the family mean of the first stage by 0.115 \ndcg{} [0.104, 0.125].

\begin{table}[!htb]
\centering
\footnotesize
\setlength{\tabcolsep}{3pt}
\resizebox{\textwidth}{!}{%
\begin{tabular}{rllccccccccrr}
\toprule
\# & System & & DL & BEIR & BRIGHT & German & Chunks & Full & Family mean & Set mean & Unanswered & \$/1k \\
\midrule
1 & Voyage Rerank 3 & R & 0.742 & 0.543 & 0.233 & 0.507 & 0.628 & 0.588 & \underline{\textbf{0.540}} & 0.450 (6) & 0.05\% & 2.84 \\
2 & Perplexity v1.2 (A) & D & 0.723 & 0.528 & 0.253 & 0.514 & 0.626 & \textbf{0.597} & \underline{\textbf{0.540}} & 0.454 (3) & 0 & 1.27 \\
3 & Cohere Rerank 4 Pro & R & 0.729 & 0.547 & 0.235 & 0.512 & 0.622 & 0.592 & \underline{\textbf{0.540}} & 0.452 (4) & 0 & 5.05 \\
4 & Perplexity v1.3 (C) & D & 0.712 & 0.529 & \textbf{0.263} & 0.530 & \textbf{0.629} & 0.568 & \underline{0.538} & \underline{\textbf{0.459}} (1) & 0 & 0.99 \\
5 & Voyage Rerank 2.5 & R & 0.730 & 0.540 & 0.235 & 0.515 & 0.621 & 0.588 & \underline{0.538} & 0.450 (5) & 0.05\% & 2.84 \\
6 & Nemotron 3.5 Rerank 8B & R & 0.718 & \textbf{0.552} & 0.205 & 0.524 & 0.627 & 0.591 & \underline{0.536} & 0.445 (11) & 0 & 2.80 \\
7 & Microsoft Decision 1 (A) & D & \textbf{0.766} & 0.522 & 0.246 & 0.487 & 0.615 & 0.576 & 0.535 & 0.444 (12) & 0 & 2.45 \\
8 & Drex Max v1.1 (A) & D & 0.716 & 0.528 & 0.242 & 0.514 & 0.621 & 0.585 & 0.534 & 0.449 (8) & 0 & 1.91 \\
9 & Liquid d1-ultra (A) & D & 0.716 & 0.529 & 0.246 & 0.510 & 0.621 & 0.583 & 0.534 & 0.449 (7) & 0 & 2.49 \\
10 & Perplexity v1.2 (C) & D & 0.701 & 0.524 & 0.257 & \textbf{0.539} & 0.625 & 0.559 & 0.534 & \underline{0.456} (2) & 0 & 0.99 \\
11 & Perplexity v1.3 (A) & D & 0.712 & 0.526 & 0.258 & 0.492 & 0.618 & 0.588 & 0.532 & 0.449 (9) & 0 & 1.25 \\
12 & Liquid d1 (A) & D & 0.724 & 0.526 & 0.245 & 0.492 & 0.614 & 0.574 & 0.529 & 0.444 (13) & 0 & 1.19 \\
13 & Jev 1.14 (A) & D & 0.710 & 0.525 & 0.250 & 0.494 & 0.616 & 0.570 & 0.528 & 0.445 (10) & 0 & 3.36 \\
14 & Morph Reflex (A) & D & 0.711 & 0.524 & 0.243 & 0.495 & 0.607 & 0.569 & 0.525 & 0.442 (14) & 0 & 2.62 \\
15 & elephant-1 (A) & D & 0.716 & 0.524 & 0.236 & 0.477 & 0.608 & 0.582 & 0.524 & 0.436 (19) & 0 & 2.24 \\
16 & StartLux v1.1 9B (A)$^{\S}$ & D & 0.707 & 0.525 & 0.246 & 0.480 & 0.611 & 0.572 & 0.524 & 0.440 (16) & 0 & 0.00 \\
17 & Cohere Rerank 4 Fast & R & 0.724 & 0.530 & 0.225 & 0.493 & 0.599 & 0.563 & 0.522 & 0.436 (20) & 0 & 4.04 \\
18 & Qwen3 Reranker 8B & R & 0.702 & 0.542 & 0.204 & 0.524 & 0.601 & 0.556 & 0.522 & 0.437 (18) & 0.10\% & 12.48 \\
19 & Mercury Decide (A) & D & 0.702 & 0.517 & 0.241 & 0.494 & 0.601 & 0.565 & 0.520 & 0.438 (17) & 0 & 2.40 \\
20 & Liquid d1.1 (C) & D & 0.697 & 0.517 & 0.247 & 0.506 & 0.602 & 0.545 & 0.519 & 0.441 (15) & 0 & 0.99 \\
21 & Drex v2.0 (A) & D & 0.695 & 0.499 & 0.243 & 0.481 & 0.598 & 0.537 & 0.509 & 0.429 (22) & 0 & 3.44 \\
22 & Perplexity Flash v1.3 (C) & D & 0.704 & 0.519 & 0.230 & 0.454 & 0.605 & 0.500 & 0.502 & 0.422 (24) & 0 & 0.50 \\
23 & GPT-6 Luna (C) & D & 0.635 & 0.498 & 0.250 & 0.527 & 0.586 & 0.484 & 0.497 & 0.434 (21) & 2.4\% & 4.66 \\
24 & Gemini Decision 1 (C) & D & 0.711 & 0.516 & 0.257 & 0.430 & 0.586 & 0.480 & 0.497 & 0.423 (23) & 6.8\% & 4.39 \\
25 & Cohere Rerank 3.5 & R & 0.705 & 0.513 & 0.150 & 0.431 & 0.584 & 0.505 & 0.481 & 0.386 (26) & 0 & 1.83 \\
26 & Perplexity Flash v1.2 (C) & D & 0.690 & 0.511 & 0.196 & 0.439 & 0.592 & 0.450 & 0.480 & 0.400 (25) & 0 & 0.50 \\
27 & Solar Decide Flash (A) & D & 0.654 & 0.483 & 0.164 & 0.444 & 0.577 & 0.533 & 0.476 & 0.385 (27) & 0 & 8.69 \\
28 & bge-reranker-v2-m3 & R & 0.702 & 0.493 & 0.127 & 0.473 & 0.554 & 0.475 & 0.471 & 0.379 (28) & 0 & 0.60 \\
29 & Drex Max v1.1 (C) & D & 0.693 & 0.384 & 0.137 & 0.494 & 0.610 & 0.410 & 0.455 & 0.363 (30) & 0 & 1.50 \\
30 & joey-1 (A) & D & 0.680 & 0.464 & 0.164 & 0.404 & 0.529 & 0.450 & 0.449 & 0.365 (29) & 0 & 0.64 \\
31 & MiniLM-L6$^{\S}$ & R & 0.701 & 0.471 & 0.121 & 0.289 & 0.497 & 0.454 & 0.422 & 0.324 (31) & 0 & 0.00 \\
\midrule
 & First stage only &  & 0.493 & 0.438 & 0.137 & 0.507 & 0.579 & 0.401 & 0.426 & 0.364 &  & 0.00 \\
\bottomrule
\end{tabular}}
\caption{Every system on all 35 sets, sorted by the family mean. \ndcg{} per family (mean of its sets), the \emph{family mean} (each of the six families weighs one sixth) and the \emph{set mean} (plain mean of the 35 sets, with the rank by this mean in parentheses). \$/1k is the cost per 1,000 queries at list price, averaged over the 35 sets. D: decision model, R: dedicated reranker. (A): one yes/no question per passage, (C): one multiple-choice question per query (Section~\ref{sec:method}). Bold: best of the column. Underlined (family mean and set mean only): level with the best of that column, within 0.004. A query the system refused, had blocked or failed scores 0. Unanswered: share of the 6,003 queries this applies to. Liquid d1.1 (C) ran the full documents and most German queries at 8,192 tokens with one question per passage (Section~\ref{sec:protocol}). Every system with a public API price is shown at that price, also where the runs used a free offer or ran self-hosted. $^{\S}$\,self-hosted, no known public price (shown at \$0, without the cost of hardware or power).}
\label{tab:overview}
\end{table}

\paragraph{Family mean.} Six systems are level at the top, within 0.004 of the best. Four of them are the dedicated rerankers Voyage Rerank 3 (0.540), Cohere Rerank 4 Pro (0.540), Voyage Rerank 2.5 (0.538) and Nemotron 3.5 Rerank 8B (0.536). The other two are Perplexity v1.2 with one question per passage (0.540) and Perplexity v1.3 with one multiple-choice question per query (0.538). Microsoft Decision 1 (0.535), Drex Max v1.1 per passage (0.534), Liquid d1-ultra (0.534), Perplexity v1.2 with multiple choice (0.534) and v1.3 per passage (0.532) follow by a small margin. Liquid d1 (0.529) and every system below it are significantly lower. Perplexity v1.2 per passage is equal to Voyage Rerank 3 and Cohere Rerank 4 Pro. It costs \$1.27 per 1,000 queries, against \$2.84 for Voyage Rerank 3 and \$5.05 for Cohere Rerank 4 Pro. Perplexity v1.3 with multiple choice costs \$0.99.

\paragraph{Set mean.} By the set mean, Perplexity v1.3 with multiple choice has the highest result (0.459). It is level with Perplexity v1.2 in the same variant. It is ahead of Perplexity v1.2 per passage, Cohere Rerank 4 Pro and both Voyage models by a small margin (0.005 to 0.009). All other systems are significantly lower.

The two means rank the systems differently because they weight the families differently. Perplexity v1.3 with multiple choice trails Voyage Rerank 3 on TREC DL (by 0.030), on BEIR (by 0.014) as well as on the full documents (by 0.021), while it leads on BRIGHT (by 0.030) and on the German sets (by 0.023). The family mean gives each family one sixth, so the three families where it trails outweigh the two where it leads. The set mean gives each set the same weight, so the 12 BRIGHT sets and the 8 German sets outweigh the 12 sets of the three families where it trails. In both means, the top systems lie within 0.01 of each other. The two best Perplexity setups are therefore as good as the best commercial rerankers tested, at 20 to 45\% of their cost on average over the 35 sets. Per family, this share ranges from 8\% to 89\% (Section~\ref{sec:cost}).

\begin{figure}[!htb]
\centering
\includegraphics[width=0.92\textwidth]{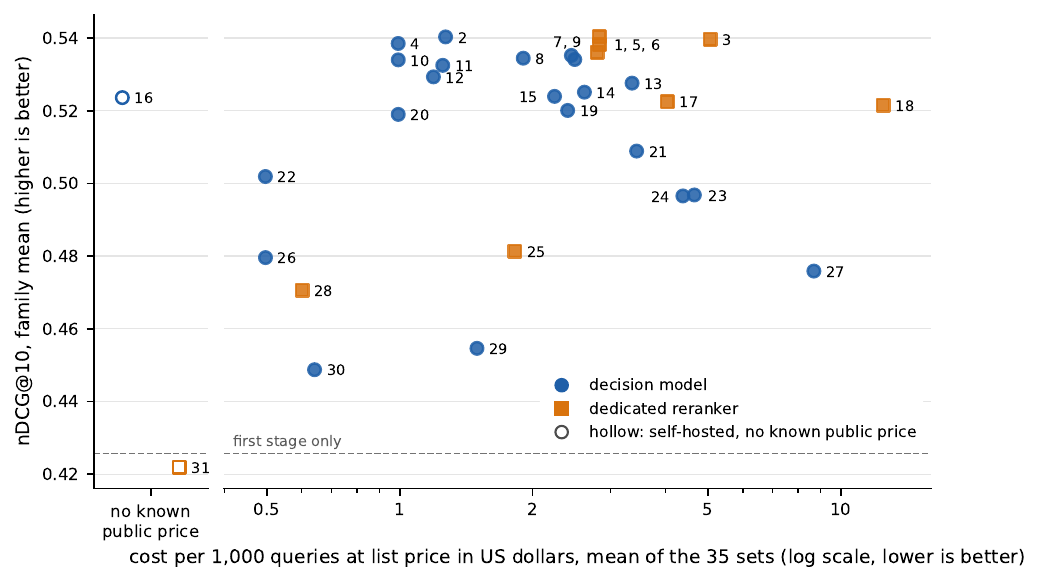}
\caption{Quality against cost for the 31 systems at public list prices. Each number is the system's number in the first column (\#) of Table~\ref{tab:overview}. The left panel shows the two self-hosted systems without a known public price. Up to \$1.27 per 1,000 queries, the best quality at each cost comes from three Perplexity setups, Perplexity Flash v1.3 (22), Perplexity v1.3 with multiple choice (4) and v1.2 per passage (2). Voyage Rerank 3 (1) is level with Perplexity v1.2 per passage at 2.2 times the cost.}
\label{fig:cost}
\end{figure}

\paragraph{Other decision models.} Figure~\ref{fig:cost} plots quality against cost. Twelve of the 22 decision-model setups, from eight vendors, rank above Cohere Rerank 4 Fast as well as Qwen3 Reranker 8B. Seven of them, from Perplexity, Microsoft, Nace and Liquid, are significantly better than both. Not every decision model reranks well. Perplexity Flash v1.2, Solar Decide Flash, Drex Max v1.1 with multiple choice and joey-1 are 0.06 to 0.09 below the top. Three more models failed the screening (Section~\ref{sec:systems}). GPT-6 Luna and Gemini Decision 1 are also low at 0.497, Gemini Decision 1 mainly because Google blocked 6.8\% of its queries (Section~\ref{sec:refusals}).

\paragraph{Perplexity v1.3 and Perplexity Flash v1.3.} Perplexity Flash v1.3 gains 0.022 [0.018, 0.027] over Perplexity Flash v1.2 on the family mean, significantly in four of the six families. Perplexity v1.3 changes only by a small margin on the family mean, up by 0.005 with multiple choice and down by 0.008 with one question per passage. With one question per passage, it is significantly lower on the long German documents ($-0.021$) and on TREC DL ($-0.011$). With multiple choice, it leans much less on the order of the candidates on SciFact (Section~\ref{sec:position}).

\paragraph{Two variants.} Perplexity v1.2 and v1.3 ran in both variants on all 35 sets. With Perplexity v1.3, multiple choice is ahead by a small margin on the family mean (0.538 against 0.532) and significantly on the set mean, by 0.011 [0.007, 0.015] (0.459 against 0.449). With Perplexity v1.2, one question per passage is ahead by a small margin on the family mean (0.540 against 0.534), while the set means are level. Drex Max v1.1 also ran in both variants. For it, one question per passage is significantly ahead, by 0.080 [0.073, 0.087] on the family mean (0.534 against 0.455). This lifts it from 29th to eighth of the 31 systems.

\paragraph{Free tier and own hardware.} Nemotron 3.5 Rerank 8B ran on OpenRouter's free tier, which has rate limits, and reaches 0.536. At DeepInfra's list price~\cite{deepinfra2026nemotron}, it would cost \$2.80 per 1,000 queries. The self-hosted StartLux v1.1 9B reaches 0.524, level with Cohere Rerank 4 Fast as well as with Qwen3 Reranker 8B.

\subsection{Short passages: TREC DL and BEIR}

On TREC DL, the highest result comes from a decision model. Microsoft Decision 1 scores 0.771 on TREC DL 2019 and 0.761 on TREC DL 2020, significantly above Cohere Rerank 4 Pro ($+0.037$ [$+0.014$, $+0.063$]) as well as Voyage Rerank 3 ($+0.024$ [$+0.002$, $+0.049$]). With only 97 queries, however, the intervals are wide. Perplexity v1.2 per passage is behind Cohere Rerank 4 Pro by a small margin (0.006) and significantly behind Voyage Rerank 3 (0.019 [0.003, 0.034]). Its multiple-choice setup scores lower (0.701). On TREC DL, even the small MiniLM-L6 reranker (22.7M parameters) keeps up with the field, since it was trained on MS MARCO, the source of TREC DL. It reaches 0.701, 24th of 31. In every other family it ranks 28th or lower.

On BEIR, the dedicated rerankers lead. Six of them score above the best decision models, five of them significantly. The sixth, Cohere Rerank 4 Fast (0.530), is level with the best decision models. The four best are Nemotron 3.5 Rerank 8B (0.552), Cohere Rerank 4 Pro (0.547), Voyage Rerank 3 (0.543) and Qwen3 Reranker 8B (0.542). The best decision-model setups are Perplexity v1.3 with multiple choice, Liquid d1-ultra (both 0.529), Drex Max v1.1 per passage and Perplexity v1.2 per passage (both 0.528). The gaps are small. Perplexity v1.2 per passage is 0.019 [0.010, 0.028] behind Cohere Rerank 4 Pro and 0.024 [0.015, 0.033] behind Nemotron 3.5 Rerank 8B.

Two sets resist reranking (Table~\ref{tab:perset-web}). On Touch\'e-2020, no system reaches BM25 (0.442), with Qwen3 Reranker 8B as the best at 0.411. This matches earlier reports. In RankGPT, every reranker stays below BM25 on this set, GPT-4 included at 0.386~\cite{sun2023rankgpt}. On Signal-1M, only six systems score above BM25, by at most 0.014 and none of them significantly.

\subsection{Reasoning-heavy retrieval: BRIGHT}

On BRIGHT, the order is reversed (Table~\ref{tab:bright}). Seventeen of the 22 decision-model setups score above the best dedicated rerankers, Voyage Rerank 2.5 and Cohere Rerank 4 Pro (both 0.235). An eighteenth, Perplexity Flash v1.3 (0.230), is behind them by a small margin, and 12 setups are significantly ahead of both. The best is Perplexity v1.3 with multiple choice (0.263), 0.028 [0.020, 0.036] above Cohere Rerank 4 Pro. Close behind are Perplexity v1.3 with one question per passage (0.258), level with the best, and Gemini Decision 1 together with Perplexity v1.2 with multiple choice (both 0.257), behind the best by a small margin. Perplexity v1.2 with multiple choice is ahead of Cohere Rerank 4 Pro by 0.022, of Voyage Rerank 3 by 0.024 and of Nemotron 3.5 Rerank 8B by 0.052, all significant. It scores higher than Cohere Rerank 4 Pro on ten of the twelve sets, significantly on five. Cohere Rerank 4 Pro is significantly ahead only on Pony. Section~\ref{sec:position} repeats some multiple-choice runs with the candidates in a random order. On the two BRIGHT sets of that control, the order explains little of this lead. Shuffling changes the result of Perplexity v1.2 with multiple choice by $-0.014$ on LeetCode and not at all on Biology. Perplexity v1.3 with multiple choice, shuffled on all twelve sets, loses only 0.006 on the BRIGHT mean (Section~\ref{sec:position}).

\begin{table}[!htb]
\centering
\scriptsize
\setlength{\tabcolsep}{2.6pt}
\resizebox{\textwidth}{!}{%
\begin{tabular}{lcccccccccccccc}
\toprule
System & Bio & Earth & Econ & Psy & Rob & Stack & Sust & Leet & Pony & AoPS & ThQ & ThT & Mean & Rank \\
\midrule
First stage only & 0.181 & 0.281 & 0.164 & 0.134 & 0.109 & 0.162 & 0.161 & 0.247 & 0.043 & 0.065 & 0.073 & 0.021 & 0.137 & -- \\
\midrule
Perplexity v1.3 (C) & 0.336 & 0.440 & \textbf{0.287} & 0.331 & 0.288 & 0.263 & 0.309 & \textbf{0.376} & 0.115 & \textbf{0.148} & \textbf{0.142} & 0.119 & \textbf{0.263} & 1 \\
Perplexity v1.3 (A) & 0.331 & 0.440 & 0.276 & 0.320 & 0.269 & 0.256 & 0.314 & 0.335 & 0.206 & 0.104 & 0.128 & 0.121 & 0.258 & 2 \\
Gemini Decision 1 (C) & 0.340 & \textbf{0.465} & 0.282 & 0.324 & 0.272 & 0.228 & \textbf{0.324} & 0.294 & 0.196 & 0.131 & 0.131 & 0.105 & 0.257 & 3 \\
Perplexity v1.2 (C) & 0.332 & 0.445 & 0.270 & \textbf{0.333} & 0.278 & 0.255 & 0.303 & 0.362 & 0.111 & 0.140 & 0.138 & 0.116 & 0.257 & 4 \\
Perplexity v1.2 (A) & 0.339 & 0.425 & 0.266 & 0.325 & 0.268 & 0.248 & 0.302 & 0.322 & 0.201 & 0.107 & 0.117 & 0.117 & 0.253 & 5 \\
GPT-6 Luna (C) & 0.315 & 0.444 & \textbf{0.287} & 0.314 & \textbf{0.291} & \textbf{0.275} & 0.300 & 0.267 & 0.139 & 0.124 & 0.132 & 0.116 & 0.250 & 6 \\
Jev 1.14 (A) & 0.319 & 0.426 & 0.279 & 0.299 & 0.253 & 0.250 & 0.300 & 0.334 & 0.192 & 0.111 & 0.118 & 0.122 & 0.250 & 7 \\
Liquid d1.1 (C) & 0.332 & 0.404 & \textbf{0.287} & 0.322 & 0.258 & 0.234 & 0.314 & 0.308 & 0.117 & 0.134 & 0.134 & 0.118 & 0.247 & 8 \\
Liquid d1-ultra (A) & 0.323 & 0.403 & 0.286 & 0.325 & 0.265 & 0.244 & 0.292 & 0.283 & 0.196 & 0.112 & 0.116 & 0.113 & 0.246 & 9 \\
Microsoft Decision 1 (A) & 0.325 & 0.409 & 0.253 & 0.307 & 0.259 & 0.239 & 0.298 & 0.340 & 0.179 & 0.119 & 0.120 & 0.109 & 0.246 & 10 \\
StartLux v1.1 9B (A) & 0.336 & 0.445 & 0.264 & 0.311 & 0.258 & 0.245 & 0.287 & 0.298 & 0.179 & 0.085 & 0.126 & 0.116 & 0.246 & 11 \\
Drex Max v1.1 (A) & \textbf{0.341} & 0.409 & 0.249 & 0.317 & 0.250 & 0.232 & 0.288 & 0.278 & \textbf{0.214} & 0.097 & 0.119 & 0.112 & 0.242 & 15 \\
Perplexity Flash v1.3 (C) & 0.328 & 0.361 & 0.272 & 0.294 & 0.253 & 0.229 & 0.286 & 0.265 & 0.111 & 0.121 & 0.121 & 0.119 & 0.230 & 21 \\
Perplexity Flash v1.2 (C) & 0.313 & 0.319 & 0.223 & 0.273 & 0.197 & 0.187 & 0.255 & 0.196 & 0.085 & 0.088 & 0.100 & 0.115 & 0.196 & 25 \\
Drex Max v1.1 (C) & 0.234 & 0.234 & 0.140 & 0.173 & 0.122 & 0.090 & 0.234 & 0.175 & 0.087 & 0.080 & 0.059 & 0.020 & 0.137 & 29 \\
\midrule
Voyage Rerank 2.5 & 0.302 & 0.418 & 0.245 & 0.293 & 0.250 & 0.234 & 0.269 & 0.277 & 0.200 & 0.108 & 0.118 & 0.107 & 0.235 & 18 \\
Cohere Rerank 4 Pro & 0.312 & 0.435 & 0.240 & 0.307 & 0.242 & 0.233 & 0.267 & 0.275 & 0.134 & 0.122 & 0.133 & 0.119 & 0.235 & 19 \\
Voyage Rerank 3 & 0.295 & 0.406 & 0.241 & 0.297 & 0.240 & 0.233 & 0.246 & 0.303 & 0.181 & 0.123 & 0.122 & 0.110 & 0.233 & 20 \\
Cohere Rerank 4 Fast & 0.310 & 0.402 & 0.238 & 0.277 & 0.203 & 0.221 & 0.257 & 0.269 & 0.149 & 0.129 & 0.126 & \textbf{0.124} & 0.225 & 22 \\
Nemotron 3.5 Rerank 8B & 0.218 & 0.381 & 0.208 & 0.261 & 0.239 & 0.200 & 0.212 & 0.286 & 0.107 & 0.125 & 0.117 & 0.102 & 0.205 & 23 \\
Qwen3 Reranker 8B & 0.227 & 0.355 & 0.229 & 0.260 & 0.250 & 0.200 & 0.225 & 0.288 & 0.099 & 0.108 & 0.111 & 0.094 & 0.204 & 24 \\
Cohere Rerank 3.5 & 0.183 & 0.284 & 0.145 & 0.206 & 0.144 & 0.150 & 0.140 & 0.246 & 0.052 & 0.085 & 0.087 & 0.078 & 0.150 & 28 \\
\bottomrule
\end{tabular}}
\caption{BRIGHT, reasoning-heavy retrieval: \ndcg{} per set after reranking the BM25 top 100, for the eleven best decision-model setups, both Perplexity Flash versions, both variants of Drex Max v1.1 as well as the seven hosted rerankers. Decision models above the line, dedicated rerankers below. Every system is in Appendix~\ref{app:perset}. Bold: best of the rows shown. Rank: position by the BRIGHT mean among all 31 systems. Columns: Biology, Earth Science, Economics, Psychology, Robotics, Stack Overflow, Sustainable Living, LeetCode, Pony, AoPS, TheoremQA-Questions, TheoremQA-Theorems.}
\label{tab:bright}
\end{table}

The two 8B rerankers, which are strong on BEIR, fall behind here, with 0.205 for Nemotron 3.5 Rerank 8B and 0.204 for Qwen3 Reranker 8B. The self-hosted decision model StartLux v1.1 9B reaches 0.246, 0.011 above Cohere Rerank 4 Pro, the joint best dedicated reranker.

Perplexity Flash v1.3 trails the best dedicated rerankers by a small margin, while Perplexity Flash v1.2 (0.196), Solar Decide Flash (0.164) and joey-1 (0.164) stay well below the four best dedicated rerankers (0.225 to 0.235). With multiple choice, Drex Max v1.1 does not improve on BM25 at all (0.137). Asked one question per passage, it reaches 0.242, ahead of the best dedicated rerankers by a small margin. The absolute level also stays low for every system, with 0.263 as the best result. Part of this is the first stage. On the two TheoremQA sets, the BM25 top 100 holds a relevant document for only 18 to 20\% of the queries (Section~\ref{sec:setup}).

\subsection{Long documents}

\paragraph{Full documents.} On TREC-News and Robust04 with full documents (column Full of Table~\ref{tab:overview}), Perplexity v1.2 per passage has the highest mean (0.597). It is ahead of the next eight systems by 0.005 to 0.015, significantly only of Drex Max v1.1 per passage (0.585). From Microsoft Decision 1 (0.576) down, the gap is significant. The multiple-choice setup of Perplexity v1.2 is significantly lower (0.559). Reading the whole document instead of its first 512 tokens helps every strong system on Robust04. Perplexity v1.2 per passage, for example, rises from 0.654 to 0.676. Some weak systems get worse. On Robust04, Perplexity Flash v1.2 falls from 0.551 to 0.490 and Cohere Rerank 3.5 from 0.555 to 0.526. Drex Max v1.1 with multiple choice, far behind with the 512-token cut, instead gains the most of all 31 systems (0.315 to 0.441).

\begin{table}[!htb]
\centering
\footnotesize
\setlength{\tabcolsep}{3.5pt}
\resizebox{\textwidth}{!}{%
\begin{tabular}{lcccccccccc}
\toprule
 & \multicolumn{2}{c}{BM25, sample 1} & \multicolumn{2}{c}{BM25, sample 2} & \multicolumn{2}{c}{mistral-embed} & \multicolumn{2}{c}{Qwen3 Embedding 8B} & & \\
\cmidrule(lr){2-3}\cmidrule(lr){4-5}\cmidrule(lr){6-7}\cmidrule(lr){8-9}
System & 1,024 & 8,192 & 1,024 & 8,192 & 1,024 & 8,192 & 1,024 & 8,192 & Mean & Rank \\
\midrule
First stage only & 0.564 & 0.564 & 0.655 & 0.655 & 0.257 & 0.257 & 0.550 & 0.550 & 0.507 & -- \\
\midrule
Perplexity v1.2 (C) & \textbf{0.532} & 0.642 & 0.580 & 0.693 & \textbf{0.328} & 0.357 & \textbf{0.524} & 0.653 & \textbf{0.539} & 1 \\
Perplexity v1.3 (C) & 0.513 & 0.655 & 0.558 & 0.686 & 0.321 & 0.356 & 0.505 & 0.650 & 0.530 & 2 \\
GPT-6 Luna (C) & 0.496 & 0.644 & 0.584 & 0.682 & 0.311 & 0.353 & 0.504 & 0.641 & 0.527 & 3 \\
Drex Max v1.1 (A) & 0.471 & 0.641 & 0.512 & 0.698 & 0.314 & 0.358 & 0.471 & 0.650 & 0.514 & 7 \\
Perplexity v1.2 (A) & 0.475 & 0.640 & 0.523 & 0.695 & 0.309 & 0.357 & 0.465 & 0.646 & 0.514 & 8 \\
Liquid d1-ultra (A) & 0.467 & 0.628 & 0.514 & 0.697 & 0.307 & 0.357 & 0.457 & 0.649 & 0.510 & 10 \\
Liquid d1.1 (C) & 0.460 & 0.634 & 0.530 & 0.672 & 0.315 & 0.346 & 0.464 & 0.626 & 0.506 & 12 \\
Drex Max v1.1 (C) & 0.515 & 0.561 & \textbf{0.585} & 0.600 & 0.318 & 0.326 & 0.509 & 0.539 & 0.494 & 15 \\
Perplexity v1.3 (A) & 0.432 & 0.626 & 0.470 & 0.686 & 0.299 & 0.357 & 0.431 & 0.638 & 0.492 & 18 \\
StartLux v1.1 9B (A) & 0.428 & 0.613 & 0.460 & 0.648 & 0.303 & 0.348 & 0.433 & 0.610 & 0.480 & 22 \\
Perplexity Flash v1.3 (C) & 0.438 & 0.503 & 0.478 & 0.568 & 0.305 & 0.345 & 0.437 & 0.559 & 0.454 & 25 \\
Perplexity Flash v1.2 (C) & 0.441 & 0.466 & 0.497 & 0.523 & 0.300 & 0.334 & 0.447 & 0.505 & 0.439 & 27 \\
Gemini Decision 1 (C) & 0.427 & 0.373 & 0.524 & 0.564 & 0.291 & 0.266 & 0.471 & 0.524 & 0.430 & 29 \\
\midrule
Qwen3 Reranker 8B & 0.449 & 0.664 & 0.541 & \textbf{0.721} & 0.324 & 0.361 & 0.470 & 0.662 & 0.524 & 4 \\
Nemotron 3.5 Rerank 8B & 0.495 & 0.650 & 0.553 & 0.686 & 0.321 & 0.357 & 0.488 & 0.638 & 0.524 & 5 \\
Voyage Rerank 2.5 & 0.441 & \textbf{0.670} & 0.510 & 0.712 & 0.307 & \textbf{0.364} & 0.454 & \textbf{0.664} & 0.515 & 6 \\
Cohere Rerank 4 Pro & 0.469 & 0.639 & 0.526 & 0.677 & 0.316 & 0.354 & 0.479 & 0.639 & 0.512 & 9 \\
Voyage Rerank 3 & 0.430 & 0.663 & 0.483 & 0.697 & 0.313 & 0.360 & 0.455 & 0.656 & 0.507 & 11 \\
bge-reranker-v2-m3 & 0.389 & 0.584 & 0.455 & 0.664 & 0.311 & 0.342 & 0.447 & 0.592 & 0.473 & 24 \\
\midrule
Google's semantic ranker, full documents$^{\ast}$ & -- & 0.514 & -- & 0.579 & -- & 0.320 & -- & 0.533 & -- & -- \\
\bottomrule
\end{tabular}}
\caption{Long German documents (MLDR): \ndcg{} after reranking the top 20 documents of four first stages, with every document cut to its first 1,024 or 8,192 tokens. BM25 samples have 100 queries each, the two embedding models 400 queries. Shown are 19 of the 31 systems, decision models above the line, dedicated rerankers below. Table~\ref{tab:perset-long} lists all 31. Bold: best reranked result of the rows shown. Rank: position by the mean over the eight sets among all 31 systems. $^{\ast}$Google's semantic ranker received the full documents instead of a cut and is shown in the 8,192 columns. According to its documentation, it reads 1,024 tokens per document, processes longer documents in segments and, in extreme cases, cuts off part of a document. It has no mean or rank. Liquid d1.1 ran most queries at 8,192 tokens with one question per passage (Section~\ref{sec:protocol}).}
\label{tab:german}
\end{table}

\paragraph{Long German documents.} Table~\ref{tab:german} shows the eight German sets. With the documents cut to 8,192 tokens, reranking improves even a strong first stage. On the Qwen3 Embedding 8B set (0.550 without reranking), 23 of the 31 systems are significantly better, and on BM25 sample 1 (0.564), 22 are. Only BM25 sample 2, the strongest first stage (0.655), is hard to beat, with three systems significantly better. At this length, the dedicated rerankers lead. Voyage Rerank 2.5 and Qwen3 Reranker 8B (both 0.602 over the four sets) are significantly ahead of every decision-model setup. The best of these are Perplexity v1.3 with multiple choice as well as Drex Max v1.1 per passage (both 0.587).

With the documents cut to 1,024 tokens, every system scores below a good first stage, even below BM25 alone. All 31 systems fall below the Qwen3 Embedding 8B first stage, Cohere Rerank 4 Pro by 0.071 and Voyage Rerank 2.5 by 0.096. Only the weak mistral-embed first stage (0.257) gains from almost every system. Since the answer can sit anywhere in these documents, a reranker that reads only the first 1,024 tokens should be left out when the first stage is good.

Averaged over all eight German sets, Perplexity v1.2 with multiple choice scores highest (0.539), followed by v1.3 in the same variant (0.530). This lead comes from the sets cut to 1,024 tokens. There, four multiple-choice setups fall the least below the first stage, namely Perplexity v1.2, Perplexity v1.3, Drex Max v1.1 and GPT-6 Luna. They lean on the first-stage order (Section~\ref{sec:position}). For Perplexity v1.3, the one model shuffled on these sets, the mean over the three sets with a good first stage (both BM25 samples and Qwen3 Embedding 8B) drops from 0.525 to 0.505 when shuffled. That is just below Nemotron 3.5 Rerank 8B (0.512), the best system outside the four, but still above all the others (Table~\ref{tab:shuffle-sets}). Gemini Decision 1 gains nothing from the longer cut, because Google blocked more of its queries at 8,192 tokens (up to 39\% of a set). Its German mean (0.430) stays below the mean of the first stages alone (0.507).

Several systems are weak on long documents. Perplexity Flash v1.2, Cohere Rerank 3.5 and Drex Max v1.1 with multiple choice gain little from the longer cut. At 8,192 tokens, they do not beat either BM25 first stage, and neither do Perplexity Flash v1.3 or joey-1.

Cohere's documentation contradicts itself on long documents. Its model overview gives Cohere Rerank 3.5 an input limit of 4k tokens and says that a longer document is split into chunks, which are all processed~\cite{cohere2026models}. Its API reference says instead that a longer document is cut to 4,096 tokens by default~\cite{cohere2026rerankapi}. To find out which holds, a German document of about 7,000 tokens went through OpenRouter, the route of the main runs. The test used four copies of this document. A sentence that answers the query sat near the start of the first copy, after more than 5,000 tokens in the second and after more than 6,000 tokens in the third. The fourth copy had no such sentence. With the sentence near the start, Cohere Rerank 3.5 gave the document a relevance score of 0.85. With the sentence further in, it gave 0.02, the same score as for the document without the sentence. Cohere Rerank 3.5 therefore reads only about the first half of a document cut to 8,192 tokens, and it scores no further chunks. Cohere Rerank 4 Pro as well as Cohere Rerank 4 Fast scored all three copies with the sentence far higher than the copy without it (0.62 to 0.95, against 0.14 for Cohere Rerank 4 Pro and 0.10 for Cohere Rerank 4 Fast). The cut at 4,096 tokens therefore did not apply to them. The main runs, like this test, left the API's limit on tokens per document unset. Asking explicitly for 8,192 tokens per document changed none of these findings.

\subsection{Chunks}

The chunk family comes closest to a production RAG pipeline. An embedding model selects 2,000-token chunks, and the top 20 go to the reranker (Table~\ref{tab:chunks}). Each chunk carries the label of its document, so a system earns the same credit for any chunk of a relevant document, whether or not that chunk answers the query. Nine systems lie within 0.01 of each other at the top. Five are the decision-model setups Perplexity v1.3 with multiple choice (0.629), Perplexity v1.2 per passage (0.626), Perplexity v1.2 with multiple choice (0.625), Drex Max v1.1 per passage (0.621) and Liquid d1-ultra (0.621). The other four are the dedicated rerankers Voyage Rerank 3 (0.628), Nemotron 3.5 Rerank 8B (0.627), Cohere Rerank 4 Pro (0.622) and Voyage Rerank 2.5 (0.621). Both Perplexity v1.2 setups are level with Cohere Rerank 4 Pro and Voyage Rerank 3. Perplexity v1.3 with multiple choice is level with Voyage Rerank 3 and ahead of Cohere Rerank 4 Pro by a small margin (0.006).

\begin{table}[!htb]
\centering
\footnotesize
\setlength{\tabcolsep}{3.5pt}
\begin{tabular}{lccccrl}
\toprule
System & MLDR German & TREC-News & Robust04 & Mean & \$/1k & vs.\ first stage \\
\midrule
First stage only & 0.401 & 0.658 & 0.677 & 0.579 & 0.00 &  \\
\midrule
Perplexity v1.3 (C) & 0.468 & 0.662 & 0.756 & \textbf{0.629} & 0.59 & $+$0.050 [$+$0.034, $+$0.066] \\
Voyage Rerank 3 & 0.451 & 0.673 & 0.761 & 0.628 & 1.56 & $+$0.050 [$+$0.035, $+$0.064] \\
Nemotron 3.5 Rerank 8B & 0.458 & 0.673 & 0.750 & 0.627 & 1.51 & $+$0.048 [$+$0.035, $+$0.062] \\
Perplexity v1.2 (A) & 0.443 & 0.661 & \textbf{0.774} & 0.626 & 0.63 & $+$0.047 [$+$0.032, $+$0.063] \\
Perplexity v1.2 (C) & \textbf{0.472} & 0.651 & 0.750 & 0.625 & 0.59 & $+$0.046 [$+$0.032, $+$0.060] \\
Cohere Rerank 4 Pro & 0.447 & \textbf{0.678} & 0.742 & 0.622 & 2.82 & $+$0.044 [$+$0.029, $+$0.059] \\
Voyage Rerank 2.5 & 0.448 & 0.658 & 0.757 & 0.621 & 1.56 & $+$0.042 [$+$0.028, $+$0.057] \\
Drex Max v1.1 (A) & 0.443 & 0.654 & 0.767 & 0.621 & 0.95 & $+$0.042 [$+$0.028, $+$0.057] \\
Liquid d1-ultra (A) & 0.445 & 0.646 & 0.770 & 0.621 & 1.29 & $+$0.042 [$+$0.026, $+$0.059] \\
Perplexity v1.3 (A) & 0.436 & 0.645 & 0.772 & 0.618 & 0.63 & $+$0.039 [$+$0.024, $+$0.055] \\
Jev 1.14 (A) & 0.429 & 0.649 & 0.769 & 0.616 & 1.57 & $+$0.037 [$+$0.020, $+$0.054] \\
Microsoft Decision 1 (A) & 0.429 & 0.647 & 0.768 & 0.615 & 1.28 & $+$0.036 [$+$0.022, $+$0.050] \\
StartLux v1.1 9B (A)$^{\S}$ & 0.428 & 0.662 & 0.744 & 0.611 & 0.00 & $+$0.033 [$+$0.018, $+$0.048] \\
Drex Max v1.1 (C) & 0.445 & 0.658 & 0.726 & 0.610 & 0.88 & $+$0.031 [$+$0.016, $+$0.045] \\
Perplexity Flash v1.3 (C) & 0.441 & 0.623 & 0.752 & 0.605 & 0.29 & $+$0.027 [$+$0.012, $+$0.041] \\
Qwen3 Reranker 8B & 0.448 & 0.636 & 0.719 & 0.601 & 6.53 & $+$0.022 [$+$0.009, $+$0.035] \\
Perplexity Flash v1.2 (C) & 0.430 & 0.625 & 0.721 & 0.592 & 0.29 & $+$0.014 [$-$0.001, $+$0.028] \\
Gemini Decision 1 (C) & 0.430 & 0.602 & 0.724 & 0.586 & 2.82 & $+$0.007 [$-$0.013, $+$0.025] \\
bge-reranker-v2-m3 & 0.419 & 0.604 & 0.638 & 0.554 & 0.54 & $-$0.025 [$-$0.041, $-$0.009] \\
\midrule
Google's semantic ranker & 0.390 & 0.646 & 0.697 & 0.578 & 1.00 & $-$0.001 [$-$0.015, $+$0.013] \\
\bottomrule
\end{tabular}
\caption{2,000-token chunks, the typical input of a RAG pipeline: \ndcg{} after reranking the top 20 chunks from Qwen3 Embedding 8B. \$/1k: cost per 1,000 queries, mean of the three sets. Last column: mean change against the first stage with its 95\% interval. Shown are 19 of the 31 systems. Table~\ref{tab:perset-long} lists all 31. Sorted by the mean. Google's semantic ranker is listed apart below the line. Bold: best of the rows shown. Every system with a public API price is shown at that price, also where the runs used a free offer or ran self-hosted. $^{\S}$\,self-hosted, no known public price (shown at \$0, without the cost of hardware or power).}
\label{tab:chunks}
\end{table}

However, the costs differ widely. Per 1,000 queries, the three Perplexity setups cost \$0.59 to \$0.63, Nemotron 3.5 Rerank 8B \$1.51, both Voyage models \$1.56, Cohere Rerank 4 Pro \$2.82 and Qwen3 Reranker 8B \$6.53. These three setups thus reach the quality of the best commercial rerankers tested at about a fifth of the cost of Cohere Rerank 4 Pro and 40\% of the cost of the Voyage models.

\subsection{Google's semantic ranker}
\label{sec:google}

Google's semantic ranker ran on 14 sets (Table~\ref{tab:google}). Over these sets, it improves on the first stage by 0.043 [0.031, 0.054], all of it on the five short-passage sets. On BRIGHT LeetCode it is significantly below BM25. On the chunks as well as the German sets taken together, it does not beat the first stage.

\begin{table}[!htb]
\centering
\footnotesize
\setlength{\tabcolsep}{3.5pt}
\resizebox{\textwidth}{!}{%
\begin{tabular}{lcccccrl}
\toprule
System & Short (5) & BRIGHT (2) & German (4) & Chunks (3) & Mean (14) & \$/1k & vs Google \\
\midrule
Voyage Rerank 3 & \textbf{0.707} & 0.299 & 0.594 & 0.628 & \textbf{0.600} & 3.35 & $+$0.072 [$+$0.063, $+$0.080] \\
Voyage Rerank 2.5 & 0.703 & 0.290 & \textbf{0.602} & 0.621 & 0.598 & 3.35 & $+$0.070 [$+$0.061, $+$0.079] \\
Perplexity v1.3 (C) & 0.683 & \textbf{0.356} & 0.587 & \textbf{0.629} & 0.597 & 1.17 & $+$0.069 [$+$0.060, $+$0.079] \\
Perplexity v1.2 (A) & 0.696 & 0.331 & 0.585 & 0.626 & 0.597 & 1.35 & $+$0.069 [$+$0.060, $+$0.078] \\
Microsoft Decision 1 (A) & 0.704 & 0.333 & 0.572 & 0.615 & 0.594 & 2.68 & $+$0.066 [$+$0.057, $+$0.076] \\
Drex Max v1.1 (A) & 0.696 & 0.310 & 0.587 & 0.621 & 0.593 & 2.02 & $+$0.065 [$+$0.057, $+$0.074] \\
Perplexity v1.3 (A) & 0.691 & 0.333 & 0.577 & 0.618 & 0.592 & 1.34 & $+$0.064 [$+$0.055, $+$0.073] \\
Cohere Rerank 4 Pro & 0.702 & 0.294 & 0.577 & 0.622 & 0.591 & 5.31 & $+$0.063 [$+$0.054, $+$0.072] \\
Perplexity v1.2 (C) & 0.671 & 0.347 & 0.586 & 0.625 & 0.591 & 1.17 & $+$0.063 [$+$0.054, $+$0.072] \\
Nemotron 3.5 Rerank 8B & 0.705 & 0.252 & 0.583 & 0.627 & 0.589 & 3.13 & $+$0.061 [$+$0.052, $+$0.069] \\
Qwen3 Reranker 8B & 0.688 & 0.258 & \textbf{0.602} & 0.601 & 0.583 & 14.25 & $+$0.055 [$+$0.047, $+$0.065] \\
StartLux v1.1 9B (A)$^{\S}$ & 0.684 & 0.317 & 0.555 & 0.611 & 0.579 & 0.00 & $+$0.051 [$+$0.043, $+$0.060] \\
Cohere Rerank 4 Fast & 0.691 & 0.289 & 0.562 & 0.599 & 0.577 & 4.25 & $+$0.049 [$+$0.040, $+$0.058] \\
GPT-6 Luna (C) & 0.639 & 0.291 & 0.580 & 0.586 & 0.561 & 5.49 & $+$0.033 [$+$0.024, $+$0.043] \\
Perplexity Flash v1.3 (C) & 0.683 & 0.296 & 0.494 & 0.605 & 0.557 & 0.58 & $+$0.029 [$+$0.020, $+$0.039] \\
Gemini Decision 1 (C) & 0.685 & 0.317 & 0.432 & 0.586 & 0.539 & 4.82 & $+$0.011 [$-$0.001, $+$0.023] \\
Perplexity Flash v1.2 (C) & 0.673 & 0.255 & 0.457 & 0.592 & 0.534 & 0.58 & $+$0.006 [$-$0.004, $+$0.016] \\
bge-reranker-v2-m3 & 0.655 & 0.167 & 0.546 & 0.554 & 0.532 & 0.68 & $+$0.004 [$-$0.004, $+$0.013] \\
\emph{Google's semantic ranker} & 0.678 & 0.162 & 0.487 & 0.578 & 0.528 & 1.00 &  \\
Cohere Rerank 3.5 & 0.668 & 0.215 & 0.444 & 0.584 & 0.521 & 1.59 & $-$0.007 [$-$0.016, $+$0.003] \\
Drex Max v1.1 (C) & 0.567 & 0.204 & 0.506 & 0.610 & 0.507 & 1.75 & $-$0.021 [$-$0.031, $-$0.011] \\
\midrule
First stage only & 0.519 & 0.214 & 0.507 & 0.579 & 0.485 & 0.00 & $-$0.043 [$-$0.054, $-$0.031] \\
\bottomrule
\end{tabular}}
\caption{Google's semantic ranker against 20 of the other 31 systems on the 14 sets it ran, which are five short-passage sets (TREC DL 2019, TREC DL 2020, TREC-COVID, NFCorpus, SciFact), two BRIGHT sets (Biology, LeetCode), the four German long-document sets (the other systems at 8,192 tokens) and the three chunk sets. Mean \ndcg{} per group and over all 14 sets, sorted by the mean over all 14 sets. \$/1k: cost per 1,000 queries on these sets. vs Google: the system's mean over the 14 sets minus Google's, with its 95\% interval. First stage only: the first stage that produced the candidates (BM25, or an embedding model for two of the German sets and for the chunks), without reranking. Bold: best of the rows shown. Every system with a public API price is shown at that price, also where the runs used a free offer or ran self-hosted. $^{\S}$\,self-hosted, no known public price (shown at \$0, without the cost of hardware or power).}
\label{tab:google}
\end{table}

On these 14 sets, 23 of the 31 systems are significantly better than Google's semantic ranker, and three are worse. Perplexity v1.3 with multiple choice is ahead of it by 0.069 [0.060, 0.079] at \$1.17 per 1,000 queries, against Google's list price of \$1.00~\cite{google2026pricing}. Even the small Perplexity Flash v1.3 is significantly better.

\subsection{Cost}
\label{sec:cost}

Table~\ref{tab:cost} gives the cost of one pass over the 35 test sets. All 31 systems together cost \$488.60 at list prices. Of this, \$59.88 was not charged. The parts below are rounded separately and add up to \$59.89. They are \$17.04 for Nemotron 3.5 Rerank 8B on OpenRouter's free tier, \$16.93 for Perplexity v1.3 as well as Perplexity Flash v1.3 on their free launch offer, \$20.84 on the free starting credit for Drex Max v1.1 (both setups), \$3.76 for the self-hosted bge-reranker-v2-m3 and \$1.32 for requests to Gemini Decision 1 on Google's free tier. Table~\ref{tab:cost-sets} in the appendix lists every set. The cheapest systems with a price are Perplexity Flash v1.2 and Perplexity Flash v1.3, at \$3.15 each for all 35 sets. The most expensive is Qwen3 Reranker 8B, which costs \$75.75 at its OpenRouter price.

\begin{table}[!htb]
\centering
\footnotesize
\setlength{\tabcolsep}{3.5pt}
\begin{tabular}{lrrrrrrrr}
\toprule
System & DL & BEIR & BRIGHT & German & Chunks & Full & Total & \$/1k \\
\midrule
Solar Decide Flash (A) & 0.47 & 9.77 & 11.90 & 18.12 & 3.16 & 7.12 & 50.54 & 8.69 \\
GPT-6 Luna (C) & 0.09 & 4.09 & 3.97 & 13.39 & 2.19 & 5.67 & 29.41 & 4.66 \\
Gemini Decision 1 (C) & 0.09 & 4.07 & 4.52 & 11.60 & 2.26 & 5.24 & 27.77 & 4.39 \\
Drex v2.0 (A) & 0.08 & 2.90 & 4.30 & 8.66 & 1.45 & 3.39 & 20.77 & 3.44 \\
Jev 1.14 (A) & 0.16 & 3.68 & 4.51 & 7.13 & 1.25 & 2.96 & 19.69 & 3.36 \\
Morph Reflex (A) & 0.09 & 2.51 & 3.33 & 6.18 & 1.06 & 2.53 & 15.69 & 2.62 \\
Liquid d1-ultra (A) & 0.06 & 2.12 & 3.15 & 6.16 & 1.02 & 2.39 & 14.91 & 2.49 \\
Microsoft Decision 1 (A) & 0.06 & 2.07 & 3.11 & 6.10 & 1.02 & 2.37 & 14.74 & 2.45 \\
Mercury Decide (A) & 0.07 & 2.15 & 3.10 & 5.73 & 0.97 & 2.36 & 14.37 & 2.40 \\
elephant-1 (A) & 0.08 & 2.15 & 2.98 & 5.19 & 0.89 & 1.96 & 13.26 & 2.24 \\
Drex Max v1.1 (A) & 0.06 & 1.76 & 2.47 & 4.43 & 0.76 & 1.89 & 11.37 & 1.91 \\
Drex Max v1.1 (C) & 0.03 & 1.30 & 1.35 & 4.29 & 0.71 & 1.79 & 9.47 & 1.50 \\
Perplexity v1.2 (A) & 0.04 & 1.17 & 1.65 & 2.96 & 0.51 & 1.26 & 7.58 & 1.27 \\
Perplexity v1.3 (A) & 0.04 & 1.13 & 1.61 & 2.95 & 0.50 & 1.25 & 7.48 & 1.25 \\
Liquid d1 (A) & 0.03 & 1.01 & 1.50 & 2.93 & 0.49 & 1.18 & 7.14 & 1.19 \\
Perplexity v1.2 (C) & 0.02 & 0.87 & 0.88 & 2.86 & 0.47 & 1.21 & 6.30 & 0.99 \\
Perplexity v1.3 (C) & 0.02 & 0.87 & 0.88 & 2.86 & 0.47 & 1.20 & 6.30 & 0.99 \\
Liquid d1.1 (C) & 0.02 & 0.86 & 0.86 & 2.90 & 0.47 & 1.14 & 6.24 & 0.99 \\
joey-1 (A) & 0.02 & 0.62 & 0.85 & 1.48 & 0.26 & 0.56 & 3.79 & 0.64 \\
Perplexity Flash v1.2 (C) & 0.01 & 0.43 & 0.44 & 1.43 & 0.24 & 0.60 & 3.15 & 0.50 \\
Perplexity Flash v1.3 (C) & 0.01 & 0.43 & 0.44 & 1.43 & 0.24 & 0.60 & 3.15 & 0.50 \\
StartLux v1.1 9B (A)$^{\S}$ & 0.00 & 0.00 & 0.00 & 0.00 & 0.00 & 0.00 & 0.00 & 0.00 \\
\midrule
Qwen3 Reranker 8B & 0.33 & 10.73 & 15.06 & 33.10 & 5.35 & 11.18 & 75.75 & 12.48 \\
Cohere Rerank 4 Pro & 0.24 & 6.05 & 6.92 & 10.85 & 2.15 & 3.58 & 29.79 & 5.05 \\
Cohere Rerank 4 Fast & 0.19 & 4.84 & 5.54 & 8.68 & 1.72 & 2.86 & 23.83 & 4.04 \\
Voyage Rerank 2.5 & 0.04 & 2.15 & 3.21 & 8.12 & 1.28 & 2.76 & 17.56 & 2.84 \\
Voyage Rerank 3 & 0.04 & 2.15 & 3.21 & 8.12 & 1.28 & 2.76 & 17.56 & 2.84 \\
Nemotron 3.5 Rerank 8B & 0.05 & 2.18 & 3.38 & 7.39 & 1.22 & 2.82 & 17.04 & 2.80 \\
Cohere Rerank 3.5 & 0.10 & 2.43 & 2.86 & 2.73 & 0.72 & 1.37 & 10.21 & 1.83 \\
bge-reranker-v2-m3 & 0.01 & 0.49 & 0.73 & 1.53 & 0.45 & 0.55 & 3.76 & 0.60 \\
MiniLM-L6$^{\S}$ & 0.00 & 0.00 & 0.00 & 0.00 & 0.00 & 0.00 & 0.00 & 0.00 \\
\midrule
All 31 systems & 2.55 & 76.95 & 98.69 & 199.28 & 34.56 & 76.56 & 488.60 &  \\
\bottomrule
\end{tabular}
\caption{What one pass over the 35 sets costs at API list prices, in US dollars, per system and family. Decision models are above the first line, dedicated rerankers below it, each group sorted by total cost from highest to lowest. Totals are computed before rounding. Last column: cost per 1,000 queries averaged over the 35 sets. Every system with a public API price is shown at that price, also where the runs used a free offer or ran self-hosted. $^{\S}$\,self-hosted, no known public price (shown at \$0, without the cost of hardware or power).}
\label{tab:cost}
\end{table}

How large the saving of a decision model is depends on the length of the candidates, because the providers price differently. Decision models, Voyage and Qwen3 Reranker 8B charge per token. Cohere charges per search, and this charge grows more slowly with the length of the candidates than a token price does. Against Cohere Rerank 4 Pro, Perplexity v1.2 is therefore cheapest in relative terms on short passages. Averaged over the two TREC DL sets, 1,000 queries cost \$0.20 with Perplexity v1.2 asked with multiple choice, \$0.41 with one question per passage and \$2.50 with Cohere Rerank 4 Pro, a factor of 12 or 6. On full documents, the factor is about 3 (\$3.57 or \$3.77 against \$11.02). Against the Voyage models, the factor is 1.1 to 2.8 for one question per passage and 2.3 to 3.7 for multiple choice, depending on the family.

\section{Analysis}
\label{sec:analysis}

\subsection{Shuffling the candidates}
\label{sec:position}

A multiple-choice request lists the candidates in the order the first stage returned them, labelled \texttt{p0}, \texttt{p1} and so on. A model can therefore use the position of an option as a hint, and tied scores keep the first-stage order. Language models are known to be sensitive to the position of content~\cite{liu2024lost,tang2024found} and to prefer certain answer labels~\cite{zheng2024selection}. Three sets were therefore rerun with the candidates in a random order for each query and compared with the run in first-stage order. In both runs, tied scores are put in a random order. One yes/no question per passage needs no such test, because each request holds only one candidate, so the model never sees the others or their order. Dedicated rerankers should score each candidate on its own as well. For the commercial rerankers this is an assumption, and a partial check supports it. Where the same query and document appeared in two German sets with different candidate lists, Voyage Rerank 3 returned the same score both times, while Voyage Rerank 2.5 as well as Cohere Rerank 4 Fast differed by less than 0.01.

\begin{table}[!htb]
\centering
\footnotesize
\setlength{\tabcolsep}{3.5pt}
\resizebox{\textwidth}{!}{%
\begin{tabular}{lcllcl}
\toprule
Set & First stage only & System & First-stage order & Shuffled & Change \\
\midrule
SciFact & 0.679 & Perplexity v1.2 (C) & 0.787 & 0.657 & $-$0.130 [$-$0.163, $-$0.097] \\
SciFact & 0.679 & Perplexity Flash v1.2 (C) & 0.780 & 0.748 & $-$0.032 [$-$0.057, $-$0.009] \\
SciFact & 0.679 & Perplexity v1.3 (C) & 0.802 & 0.753 & $-$0.049 [$-$0.070, $-$0.028] \\
SciFact & 0.679 & Perplexity Flash v1.3 (C) & 0.795 & 0.759 & $-$0.036 [$-$0.059, $-$0.014] \\
SciFact & 0.679 & GPT-6 Luna (C) & 0.782 & 0.793 & $+$0.011 [$-$0.008, $+$0.029] \\
SciFact & 0.679 & Gemini Decision 1 (C) & 0.800 & 0.803 & $+$0.003 [$-$0.012, $+$0.019] \\
SciFact & 0.679 & Drex Max v1.1 (C) & 0.525 & 0.180 & $-$0.345 [$-$0.398, $-$0.289] \\
\midrule
BRIGHT Biology & 0.181 & Perplexity v1.2 (C) & 0.332 & 0.332 & $-$0.000 [$-$0.011, $+$0.010] \\
BRIGHT Biology & 0.181 & Perplexity Flash v1.2 (C) & 0.313 & 0.287 & $-$0.026 [$-$0.054, $-$0.001] \\
BRIGHT Biology & 0.181 & Perplexity v1.3 (C) & 0.336 & 0.334 & $-$0.002 [$-$0.013, $+$0.009] \\
BRIGHT Biology & 0.181 & Perplexity Flash v1.3 (C) & 0.328 & 0.321 & $-$0.007 [$-$0.023, $+$0.009] \\
BRIGHT Biology & 0.181 & GPT-6 Luna (C) & 0.290 & 0.312 & $+$0.022 [$-$0.002, $+$0.046] \\
BRIGHT Biology & 0.181 & Gemini Decision 1 (C) & 0.340 & 0.337 & $-$0.004 [$-$0.022, $+$0.013] \\
\midrule
BRIGHT LeetCode & 0.247 & Perplexity v1.2 (C) & 0.362 & 0.348 & $-$0.014 [$-$0.029, $-$0.001] \\
BRIGHT LeetCode & 0.247 & Perplexity Flash v1.2 (C) & 0.196 & 0.233 & $+$0.037 [$+$0.005, $+$0.068] \\
BRIGHT LeetCode & 0.247 & Perplexity v1.3 (C) & 0.376 & 0.378 & $+$0.002 [$-$0.011, $+$0.016] \\
BRIGHT LeetCode & 0.247 & Perplexity Flash v1.3 (C) & 0.265 & 0.266 & $+$0.002 [$-$0.031, $+$0.035] \\
BRIGHT LeetCode & 0.247 & GPT-6 Luna (C) & 0.244 & 0.274 & $+$0.029 [$-$0.002, $+$0.063] \\
BRIGHT LeetCode & 0.247 & Gemini Decision 1 (C) & 0.294 & 0.290 & $-$0.004 [$-$0.024, $+$0.018] \\
\bottomrule
\end{tabular}}
\caption{Shuffle control for the multiple-choice variant with up to 100 options per request. \emph{First stage only} is the \ndcg{} of the first stage alone, without reranking. \emph{First-stage order} is the \ndcg{} of the model with the candidates listed in first-stage order, and \emph{Shuffled} is its \ndcg{} with the candidates listed in a random order. \emph{Change} is Shuffled minus First-stage order, with its 95\% interval. In both runs, tied scores are put in one random order, so the \emph{First-stage order} column differs from the main tables by at most 0.002. The exception is GPT-6 Luna, whose many ties lower its result by 0.005 on SciFact, 0.025 on BRIGHT Biology and 0.022 on BRIGHT LeetCode in this draw. Refused or blocked queries score 0 in both runs. GPT-6 Luna refused partly different queries in the two orders (on BRIGHT LeetCode 16 against 9), so its changes reflect refusals as well as the order. Grouped by set, with the models in the same order in each group.}
\label{tab:shuffle}
\end{table}

Table~\ref{tab:shuffle} shows the outcome. On SciFact, where BM25 alone is strong (0.679), Perplexity v1.2 loses 0.130 when the order is shuffled and ends below BM25. Perplexity Flash v1.2 loses 0.032, while GPT-6 Luna loses nothing. On the two BRIGHT sets, where the first-stage order carries little information, the order matters far less for Perplexity v1.2 (no change on Biology, $-0.014$ [$-0.029$, $-0.001$] on LeetCode). Perplexity Flash v1.2 loses 0.026 [0.001, 0.054] on Biology, but on LeetCode, where it stays below BM25 in either order, shuffling raises its result by 0.037 [0.005, 0.068]. Gemini Decision 1 does not depend on the order. Shuffling leaves it level on all three sets ($+0.003$, $-0.004$, $-0.004$). GPT-6 Luna even does better shuffled on the BRIGHT sets, partly because it refused fewer queries in the shuffled run (Section~\ref{sec:refusals}). Drex Max v1.1, shuffled on SciFact only, depends on the order heavily. Its result falls from 0.525 to 0.180 ($-0.345$ [$-0.398$, $-0.289$]), far below BM25. Asked one question per passage, which hides the order, it reaches 0.793 there. In the pilot, which used only 20 candidates from a separate BM25 run, the order mattered less. Shuffling changed each of the three multiple-choice models tested there by at most 0.016 on SciFact.

On SciFact, Perplexity v1.3 leans much less on the order than v1.2. It scores higher than Perplexity v1.2 with the first-stage order (0.802 against 0.787), and it loses less than half as much when the order is shuffled (0.049 [0.028, 0.070] against 0.130). Shuffled, Perplexity v1.3 is thus 0.096 ahead of v1.2 (0.753 against 0.657), so it copes much better with 100 options in a random order. With one question per passage, the two versions are level there (0.799 against 0.800). On the two BRIGHT sets it loses nothing ($-0.002$ and $+0.002$). Perplexity Flash v1.3 shows no such progress. On SciFact it loses 0.036 [0.014, 0.059], about as much as Perplexity Flash v1.2 (0.032).

\begin{table}[!htb]
\centering
\footnotesize
\setlength{\tabcolsep}{3.5pt}
\begin{tabular}{lcccl}
\toprule
Family & First stage only & First-stage order & Shuffled & Change \\
\midrule
TREC DL & 0.493 & 0.712 & 0.686 & $-$0.026 [$-$0.039, $-$0.013] \\
BEIR & 0.438 & 0.529 & 0.495 & $-$0.034 [$-$0.041, $-$0.028] \\
BRIGHT & 0.137 & 0.263 & 0.257 & $-$0.006 [$-$0.010, $-$0.002] \\
German & 0.507 & 0.530 & 0.522 & $-$0.009 [$-$0.014, $-$0.003] \\
Chunks & 0.579 & 0.629 & 0.622 & $-$0.007 [$-$0.015, $+$0.000] \\
Full documents & 0.401 & 0.568 & 0.539 & $-$0.029 [$-$0.042, $-$0.016] \\
\midrule
Family mean & 0.426 & 0.538 & 0.520 & $-$0.018 [$-$0.022, $-$0.015] \\
Set mean & 0.364 & 0.459 & 0.444 & $-$0.015 [$-$0.018, $-$0.013] \\
\bottomrule
\end{tabular}
\caption{Perplexity v1.3 (C) on all 35 sets, once with the candidates labelled in first-stage order and once in a random order for each query. Mean \ndcg{} per family as well as overall, plus the change with its 95\% interval (paired bootstrap, stratified by set). On the chunks, the interval reaches $+$0.0002, so that change is not significant. Ties are broken randomly on both sides. Every set is in Table~\ref{tab:shuffle-sets} in the appendix. The families follow the column order of Table~\ref{tab:overview}.}
\label{tab:shuffle-all}
\end{table}

To see how much of the main result depends on the order, Perplexity v1.3 also ran with multiple choice and shuffled candidates on all 35 sets (Table~\ref{tab:shuffle-all}). Shuffling lowers its family mean from 0.538 to 0.520, a loss of 0.018 [0.015, 0.022], and its set mean by 0.015 [0.013, 0.018]. The loss is largest on TREC DL, BEIR and the full documents, where each request holds up to 100 candidates ($-0.026$, $-0.034$, $-0.029$). It is small on BRIGHT ($-0.006$), whose first-stage order carries little information, as well as on the German and chunk sets with 20 candidates ($-0.009$, $-0.007$), although these two families have the strongest first stages. The German mean hides a split, however. On the three sets cut to 1,024 tokens with a good first stage, Perplexity v1.3 loses 0.019 to 0.022, close to its loss on TREC DL, while no set at 8,192 tokens changes significantly (Table~\ref{tab:shuffle-sets}). Shuffled, the multiple-choice setup falls below the family mean of Perplexity v1.3 with one question per passage (0.532), whose scores do not depend on the order.

With about 100 options, Perplexity v1.2 thus relies on the candidate order on SciFact, where BM25 already puts most relevant candidates near the top. Asked one question per passage, which hides the order, it reaches 0.800 on SciFact, so it can judge the candidates well without the order. Part of a multiple-choice result therefore comes from the first stage. For Perplexity v1.3, part of the multiple-choice result comes from the order on the sets with 100 candidates outside BRIGHT, whether the first stage is strong or weak. Shuffled, it loses 0.049 on SciFact (BM25 0.679) but also 0.042 on NFCorpus (BM25 0.337, Table~\ref{tab:shuffle-sets}). On the two BRIGHT sets, Perplexity v1.2 loses at most 0.014 when shuffled, against its lead of 0.059 [0.029, 0.092] over the best dedicated reranker on LeetCode. On Biology, its lead is 0.020 and not significant. In a pipeline, leaning on the order of a good first stage does no harm, since that order is mostly right. It becomes a risk when the order of the candidates carries no information, for example in a hybrid search that merges two candidate lists without a merged ranking. One yes/no question per passage is safer in that case, as long as the model does not round its scores.

\paragraph{Ties.} Some models produce scores that often cannot separate the candidates. GPT-6 Luna rounds its probabilities. On SciFact and BRIGHT Biology, it gives only 4 or 5 different scores to the 100 candidates of a query, so most candidates share a score. On SciFact, its top ten equal the top ten of the first stage for 32\% of the queries. Breaking its ties randomly instead of by the first-stage order lowers its result, on average over all tie orders, from 0.315 to 0.293 on BRIGHT Biology and from 0.267 to 0.249 on LeetCode. Jev 1.14 rounds its probabilities to two decimals with one question per passage, so the first-stage order also decides part of its ranking. On Touch\'e-2020, where BM25 beats every system, breaking its ties randomly lowers its result from 0.346 to 0.279. On the eight other measured sets, the change is at most 0.012. Drex Max v1.1, in contrast, spreads its probability thinly. The median of its highest option probability is 0.107 on SciFact, against 0.861 for Perplexity v1.2 and 0.806 for Perplexity Flash v1.2. Tied candidates should therefore keep the first-stage order, as the drops for GPT-6 Luna and Jev 1.14 show.

\subsection{Request size}
\label{sec:size}

Multiple choice puts all candidates into one request, so a request with 100 passages of up to 512 tokens can exceed 60,000 tokens. Table~\ref{tab:size} sorts the queries of the 22 sets of TREC DL, BEIR and BRIGHT into four groups of equal size by the length of their request. For each group, it gives the mean gain of each model over BM25. Each set has requests of a typical length, so the groups also differ in which sets they hold. The smallest group is about half DBPedia, the second almost all BRIGHT. Perplexity v1.2, Perplexity v1.3, Liquid d1.1 and Gemini Decision 1 gain at least as much on the largest requests as on the smallest (v1.3 $+0.110$ in the smallest group against $+0.133$ in the largest). GPT-6 Luna gains less only on the smallest requests ($+0.079$, against $+0.104$ to $+0.114$ in the other groups). Perplexity Flash v1.2 gains less beyond 13,000 tokens but stays clearly positive, while Perplexity Flash v1.3 holds its gain at every size ($+0.092$ to $+0.104$). Drex Max v1.1 is the exception. It gains 0.112 on the smallest requests, as much as the best models, but only 0.027 in the second group. In the two upper groups, it loses 0.102 and 0.085 against BM25.

\begin{table}[!htb]
\centering
\footnotesize
\setlength{\tabcolsep}{3.5pt}
\begin{tabular}{lcccc}
\toprule
System & 0.2k--13.3k & 13.4k--31.6k & 31.7k--41.8k & 41.8k--67.2k \\
\midrule
Perplexity v1.3 (C) & $+$0.110 & $+$0.117 & $+$0.123 & $+$0.133 \\
Perplexity v1.2 (C) & $+$0.110 & $+$0.113 & $+$0.111 & $+$0.119 \\
Liquid d1.1 (C) & $+$0.092 & $+$0.105 & $+$0.115 & $+$0.106 \\
GPT-6 Luna (C) & $+$0.079 & $+$0.114 & $+$0.104 & $+$0.107 \\
Gemini Decision 1 (C) & $+$0.117 & $+$0.125 & $+$0.115 & $+$0.124 \\
Perplexity Flash v1.3 (C) & $+$0.100 & $+$0.092 & $+$0.104 & $+$0.104 \\
Perplexity Flash v1.2 (C) & $+$0.100 & $+$0.062 & $+$0.082 & $+$0.067 \\
Drex Max v1.1 (C) & $+$0.112 & $+$0.027 & $-$0.102 & $-$0.085 \\
\bottomrule
\end{tabular}
\caption{Request size in the multiple-choice variant (up to 100 options per request). For each model, the queries of the 22 sets of TREC DL, BEIR and BRIGHT are split into four groups of equal size by the input tokens of their request, from the smallest requests on the left to the largest on the right. Each cell is the mean change in \ndcg{} against BM25 over the queries in that group. Unlike elsewhere, every query counts the same, so sets with many queries weigh more. The column heads give the bounds of each group for Perplexity v1.2, rounded to 100 tokens. The other providers count tokens slightly differently. Refused or blocked queries are left out, as are queries that fell back to one question per passage. Unlike in the other tables, these changes have no interval. Sorted by the set mean of Table~\ref{tab:overview}.}
\label{tab:size}
\end{table}

Per set, Drex Max v1.1 shows the same pattern. It gains on TREC DL and DBPedia, where a request has about 10,000 tokens. It loses on most BEIR sets, with 35,000 to 53,000 tokens per request. On the German sets, which have 20 candidates per query, it stays within 0.02 of Perplexity v1.2 when the documents are cut to 1,024 tokens (about 20,000 tokens per request). At 8,192 tokens it falls 0.08 to 0.11 behind on the three sets whose requests reach about 160,000 tokens. Request size alone does not explain the pattern, though. On the full documents, with requests of 143,000 to 162,000 tokens, it does better than with the 512-token cut. These runs cannot separate the number of options, the size of the request and the content. One question per passage keeps most requests small. The per-passage setup of Drex Max v1.1 is ahead of its multiple-choice setup by 0.144 on BEIR, 0.105 on BRIGHT and 0.175 on the full documents, but by only 0.023 on TREC DL.

\subsection{Prompts}
\label{sec:prompt}

The two prompts of Section~\ref{sec:method} were written once and never tuned. To see what tuning could add, five other prompts for the multiple-choice question were tested on seven sets with Perplexity v1.2 and Perplexity Flash v1.2 (Table~\ref{tab:prompt} in the appendix). The best prompt was chosen on a fixed random half of the queries, and all prompts are reported on the other half. The best prompt for Perplexity v1.2 beats the default prompt by a small margin (0.009), so the method works without prompt engineering.

\subsection{Refusals, blocked requests and failures}
\label{sec:refusals}

Two decision models do not answer every query. None of these refusals should happen, since the test data come from public, harmless and well-known benchmarks. GPT-6 Luna refused 142 of the 6,003 queries (2.4\%, Table~\ref{tab:refusals}). Its refusals concentrate on a few sets, with 17.5\% of the queries of BRIGHT TheoremQA-Questions, 11.3\% of LeetCode, 10.5\% of TheoremQA-Theorems and 9.3\% of Signal-1M, which matches news headlines against tweets. Refusals were final. In the pilot they repeated on every retry, and in the main runs each refused query \emph{was refused again on all four retries}. Which queries are refused still depends on the order of the candidates. Shuffled, GPT-6 Luna refused 9 instead of 16 queries of LeetCode and partly different ones on BRIGHT Biology. Through OpenRouter, one refused question fails the whole request, which means that a refused multiple-choice request returns no score for any candidate.

\begin{table}[!hb]
\centering
\footnotesize
\setlength{\tabcolsep}{3.5pt}
\begin{tabular}{lrrr}
\toprule
Set & Queries & GPT-6 Luna refused & Gemini Decision 1 blocked \\
\midrule
BM25 sample 1, 8,192 tokens & 100 & 1 (1.0\%) & 39 (39\%) \\
Touch\'e-2020 & 49 & 0 & 17 (35\%) \\
mistral-embed, 8,192 tokens & 400 & 3 (0.8\%) & 110 (28\%) \\
BM25 sample 2, 8,192 tokens & 100 & 0 & 19 (19\%) \\
BRIGHT TheoremQA-Questions & 194 & 34 (18\%) & 0 \\
Qwen3 Embedding 8B, 8,192 tokens & 400 & 2 (0.5\%) & 63 (16\%) \\
Robust04, full documents & 249 & 1 (0.4\%) & 39 (16\%) \\
TREC-News, full documents & 57 & 2 (3.5\%) & 7 (12\%) \\
BRIGHT LeetCode & 142 & 16 (11\%) & 0 \\
BM25 sample 1, 1,024 tokens & 100 & 4 (4.0\%) & 11 (11\%) \\
BRIGHT TheoremQA-Theorems & 76 & 8 (11\%) & 0 \\
Signal-1M & 97 & 9 (9.3\%) & 0 \\
mistral-embed, 1,024 tokens & 400 & 9 (2.2\%) & 26 (6.5\%) \\
\midrule
All 35 sets & 6,003 & 142 (2.4\%) & 409 (6.8\%) \\
\bottomrule
\end{tabular}
\caption{Queries refused by GPT-6 Luna or blocked by Google for Gemini Decision 1, on every set where either model loses more than 6\% of its queries, plus the totals over all 35 sets. Sorted by the larger of the two shares, highest first. Such a query scores 0 in every table except Table~\ref{tab:size}, which leaves it out. No other decision model refused or had a query blocked.}
\label{tab:refusals}
\end{table}

For Gemini Decision 1, Google blocked the request of 409 of the 6,003 queries (6.8\%) before the model saw it. The error message states that ``the prompt contains sensitive words that violate Google's Generative AI Prohibited Use policy''. The blocks follow the content of the candidates. There were none on TREC DL or on five of the eight BEIR sets, and only two on BRIGHT. In contrast, \textbf{39\%} of the queries were blocked on BM25 sample 1 at 8,192 tokens, \textbf{28\%} on the mistral-embed set at 8,192 tokens and \textbf{35\%} (17 of 49) on Touch\'e-2020, whose queries are ordinary debate topics. A single candidate is often enough. One German forum page appears in 61 blocked queries of the mistral-embed set but in no answered one.

A blocked request returns an HTTP error without scores, and retrying does not help. In an integration that does not catch this case, every blocked query either shows the user an error or leaves the candidates unranked. Under the rule of Section~\ref{sec:protocol}, such a query scores 0, which lowers the family mean of Gemini Decision 1 to 0.497. If the first-stage order were kept for the blocked queries, it would be 0.527. To see what an integration can recover, every blocked query was sent again with one question per passage. In 185 of the 409 queries, exactly one candidate was blocked on its own, in 125 several were. In the remaining 99, no single candidate was blocked, so only the combination of the candidates triggered the filter. An integration that falls back to one question per passage after a block, ranking the blocked candidates last, reaches a family mean of 0.530. This takes one request per candidate for each blocked query. No other decision model refused a query or had one blocked.

Permanent technical failures were rare. The two Voyage models lost three queries each to their limit of 600,000 tokens per request, two of BRIGHT Robotics and one of Robust04 with full documents. Qwen3 Reranker 8B lost six queries of Robust04 with full documents, two because they exceeded the input limit of the served model (40,959 tokens), four to an overloaded service. GPT-6 Luna lost one query. With one question per passage, a single candidate sometimes came back without a score, so it was ranked last. This happened to 42 candidates for Liquid d1.1, 38 for Drex Max v1.1 with multiple choice, 28 for Drex Max v1.1 per passage, 17 for Gemini Decision 1 (blocked), plus one each for joey-1 and elephant-1.

\subsection{A high index position gives no assurance of reranking quality}
\label{sec:aa}

The Artificial Analysis Decision Index~\cite{aa2026decisionindex} is a public leaderboard for decision models. It combines benchmarks of general decision making but contains no retrieval or ranking task. On 13 November 2026, fourteen models were both on the index and in the main runs (Table~\ref{tab:aa}). For the three models that ran in both variants, the table uses multiple choice, the cheaper variant that needs one request per query. The rank correlation between the index and the family mean is not significant (Spearman $+0.30$, $p=0.30$, or $+0.34$, $p=0.24$, with the set mean). With fourteen models, it is too imprecise to call weak or strong. \textbf{The top decision model of the index on that day, Drex Max v1.1, is the weakest of the fourteen as a reranker.} It is last of the fourteen on three of the six families, namely BEIR, BRIGHT and the full documents. On large requests, it often \textbf{falls below BM25} (Section~\ref{sec:size}). Microsoft Decision 1, in contrast, is ninth of the fourteen on the index but second as a reranker by the family mean (fifth by the set mean).

\begin{table}[!htb]
\centering
\footnotesize
\setlength{\tabcolsep}{3.5pt}
\begin{tabular}{lccccc}
\toprule
System & Index score & Index rank & Reranker rank & Family mean & Set mean \\
\midrule
Drex Max v1.1 (C) & 72.6 & 1 & 14 & 0.455 & 0.363 \\
Liquid d1-ultra (A) & 70.6 & 2 & 3 & 0.534 & 0.449 \\
Perplexity v1.3 (C) & 70.4 & 3 & 1 & 0.538 & 0.459 \\
Perplexity v1.2 (C) & 69.8 & 4 & 4 & 0.534 & 0.456 \\
elephant-1 (A) & 66.4 & 5 & 7 & 0.524 & 0.436 \\
Jev 1.14 (A) & 66.2 & 6 & 6 & 0.528 & 0.445 \\
Gemini Decision 1 (C) & 66.0 & 7 & 12 & 0.497 & 0.423 \\
Perplexity Flash v1.3 (C) & 64.1 & 8 & 10 & 0.502 & 0.422 \\
Microsoft Decision 1 (A) & 62.1 & 9 & 2 & 0.535 & 0.444 \\
Mercury Decide (A) & 62.1 & 10 & 8 & 0.520 & 0.438 \\
Drex v2.0 (A) & 61.6 & 11 & 9 & 0.509 & 0.429 \\
Liquid d1 (A) & 60.8 & 12 & 5 & 0.529 & 0.444 \\
GPT-6 Luna (C) & 60.2 & 13 & 11 & 0.497 & 0.434 \\
Perplexity Flash v1.2 (C) & 60.1 & 14 & 13 & 0.480 & 0.400 \\
\midrule
Drex Max v1.1 (A) & 72.6 & 1 & 3 & 0.534 & 0.449 \\
\bottomrule
\end{tabular}
\caption{The Artificial Analysis Decision Index (version 2.1, as read on 13 November 2026) against reranking quality for the fourteen decision models that were both on the index and in the main runs. (A) marks one yes/no question per passage, (C) one multiple-choice question per query. Perplexity v1.2 and v1.3 as well as Drex Max v1.1 ran in both variants and are listed with multiple choice, the cheaper variant that needs one request per query. Index rank: rank among these fourteen by index score. Reranker rank: rank among these fourteen by family mean. Below the line, Drex Max v1.1 with one question per passage, ranked as if it took the place of its multiple-choice row. elephant-1 appears on the index as Elephant 9B. Refused or blocked queries score 0.}
\label{tab:aa}
\end{table}

Asked one question per passage instead, Drex Max v1.1 is third of the fourteen, 0.080 [0.073, 0.087] above its multiple-choice setup on the family mean. This run was added only after the multiple-choice result was known. With it in the table, the correlation is $+0.66$ ($p=0.012$) for the family mean and $+0.71$ ($p=0.005$) for the set mean. One question per passage costs Drex Max v1.1 about 27\% more, however, and needs one request per candidate. A high position on the index thus gives no assurance for reranking. A model that is to be used as a reranker has to be tested on retrieval data.

If the index added a reranking test, it should ask with multiple choice, which costs less and needs only one request per query. The best decision models on the 35 sets, Perplexity v1.2 and v1.3, gain little or nothing from one question per passage (Section~\ref{sec:results}).

Later on 13 November, the index added Liquid d1.1, which is also in the main runs, in fifth place among these models. With Liquid d1.1, the correlation is still not significant ($+0.30$ for the family mean, $+0.36$ for the set mean), while Drex Max v1.1 remains the weakest reranker.

\subsection{Self-hosting}
\label{sec:selfhost}

Eight of the 19 decision models have open weights (Table~\ref{tab:systems}), so they can run without an API. This was tried on one consumer graphics card with 24\,GB. The card ran under Windows while also driving the displays, with very little tuning, so the speeds below are certainly a lower bound.

StartLux v1.1 9B ran all 35 sets on this card. Its family mean of 0.524 is level with Cohere Rerank 4 Fast as well as Qwen3 Reranker 8B. It is 0.016 [0.011, 0.021] below Cohere Rerank 4 Pro but 0.044 [0.037, 0.051] above Perplexity Flash v1.2. The card processed about 11 candidates per second on TREC DL passages, 3 to 8 on the BRIGHT sets and 1.4 to 2.2 on 2,000-token chunks, so a query with 100 short passages takes about nine seconds. That is enough for batch work but too slow for interactive search. In addition, its licence allows non-commercial use only.

Perplexity Flash v1.2 (4.54B parameters, Apache 2.0) ran on TREC DL 2019 with the vendor's reference code. With multiple choice, it reproduces the API result (0.696 against 0.695) at a median of 3.6 seconds per query, using 12.3\,GB of memory. With one question per passage, scored by the difference between the logits of yes and no, it reaches 0.726 at 5.4 seconds per query. This gain over multiple choice was not tested for significance. Perplexity's package runs fast only on NVIDIA cards, and the card used here is from AMD, so these speeds are certainly a lower bound.

\subsection{Latency}
\label{sec:latency}

Latency was measured from one client in Europe, without control over the load of the providers or the time of day, so the numbers are only an indication. For systems that need one request per query, the median on TREC DL 2019 (100 short passages) is 0.4 to 0.9 seconds for six of the eight multiple-choice decision models. Gemini Decision 1 (0.41), Perplexity Flash v1.2 (0.42) and GPT-6 Luna (0.42) are the quickest of these. Perplexity v1.2 (0.61), Liquid d1.1 (0.63) and Drex Max v1.1 (0.87) take longer. The commercial rerankers take 0.4 to 0.8 seconds, from Voyage Rerank 2.5 (0.42) to Cohere Rerank 4 Fast (0.79). Qwen3 Reranker 8B (1.71) and Nemotron 3.5 Rerank 8B on the free tier (1.95) are slower. On their free launch offer, Perplexity Flash v1.3 took 0.39 seconds and Perplexity v1.3 took 1.20 seconds.

Large requests change this picture. On the Qwen3 Embedding 8B set at 8,192 tokens, with 20 documents in one request of about 160,000 tokens, Perplexity v1.2 takes 7.1 seconds and Perplexity Flash v1.2 7.4. On the free launch offer, Perplexity v1.3 takes 9.0 seconds, Perplexity Flash v1.3 8.6. Gemini Decision 1 (1.6 seconds), Cohere Rerank 4 Pro (2.4) and Voyage Rerank 3 (1.0) are much faster, while Qwen3 Reranker 8B takes 9.3 seconds. On the two BM25 sets of the same size, Perplexity Flash v1.3 took 14.4 and 18.4 seconds. One question per passage avoids the large request by sending the candidates in parallel, so its latency is that of the slowest of these requests. In the pilot, the slowest of 20 requests with 4,096 tokens each took a median of 1.0 seconds, where the single multiple-choice request took 9.7. In the main runs, the slowest request of a query took a median of 1.1 seconds for Perplexity v1.2 on the Qwen3 Embedding 8B set, against 7.1 seconds for the single multiple-choice request. On TREC DL 2019, with 100 short passages, it was the other way round, with 1.8 seconds against 0.6. These per-passage times assume that all requests of a query are sent at the same moment. The runs sent at most eight requests at a time (twelve for Perplexity v1.3), so a query with 100 candidates went out in about 13 rounds (9 for v1.3) and took longer.

\section{Discussion}
\label{sec:discussion}

\subsection{Choosing a system}

Which system is best depends on the workload. The results give a clear set of rules.

\begin{itemize}
\item \textbf{Chunks and mixed workloads.} The best Perplexity setups and the best commercial rerankers tested are level in quality (on chunks, judged by the labels of whole documents). Since these setups cost only 20 to 45\% as much, they are the better choice.
\item \textbf{Reasoning-heavy queries.} On BRIGHT, good decision models are ahead of every dedicated reranker tested, so a decision model is the better choice there.
\item \textbf{Short passages.} On BEIR, dedicated rerankers keep a lead of about 0.02 \ndcg{}, while on TREC DL a decision model, Microsoft Decision 1, scores highest. Whether the lead on BEIR is worth the higher cost depends on the application. At DeepInfra's list price, Nemotron 3.5 Rerank 8B offers it at about 40\% of the cost of Cohere Rerank 4 Pro on BEIR (\$1.64 against \$4.27 per 1,000 queries). Perplexity v1.3 with multiple choice, one of the two best decision-model setups there, costs \$0.65.
\item \textbf{Long documents.} When the answer can sit anywhere in a document, the reranker must read far into it. On MLDR German, whose queries were generated from paragraphs anywhere in a document, a reranker that read only the first 1,024 tokens scored below a good first stage, even below BM25 alone. At 8,192 tokens, two dedicated rerankers, Voyage Rerank 2.5 and Qwen3 Reranker 8B, scored highest, about 0.015 above the best decision models. Voyage Rerank 2.5 costs about three times as much there as Perplexity v1.3 with multiple choice, so the choice comes down to cost.
\item \textbf{No API budget or data sovereignty.} Several models whose open weights allow commercial use stay within 0.02 of the top family mean. They are Perplexity v1.2, Perplexity v1.3 (both Apache 2.0, 26.1B), elephant-1 (MIT, 9.4B), Nemotron 3.5 Rerank 8B (OpenMDW) and Qwen3 Reranker 8B (Apache 2.0). Whoever has the weights can run the model on their own hardware, so the data stay in their hands. A decision model can also serve other tasks, such as classification, so one self-hosted model can cover several needs where a dedicated reranker only reranks. These models ran only through APIs, so their speed on local hardware was not measured. StartLux v1.1 9B, which ran all 35 sets on one consumer graphics card, also stays within 0.02, but its licence allows only non-commercial use, and the card needs about nine seconds for a query with 100 short passages (Section~\ref{sec:selfhost}). Nemotron 3.5 Rerank 8B also runs on OpenRouter's free tier, but with tight rate limits, and only if the provider may train on the submitted data, which rules out confidential documents.
\item \textbf{Filtered or refused content.} A model whose provider filters the input, or that refuses questions, can fail whole queries. Google blocked up to \textbf{39\%} of the queries to Gemini Decision 1 on long German documents, as well as more than a third on a set of debate queries, in both cases because of the candidates. GPT-6 Luna refused \textbf{18\%} of a set of maths and science queries. With such a model, an integration needs a fallback for blocked or refused requests, which adds latency, or users will see errors. Even a refusal rate of 1\% means that 1\% of all queries fail. For Gemini Decision 1, falling back to one question per passage recovers about as much as keeping the first-stage order (family mean 0.530 against 0.527).
\end{itemize}

\paragraph{Which variant.} For Perplexity v1.2 and v1.3, which ran in both variants on all 35 sets, multiple choice is the better default. With the current Perplexity v1.3, multiple choice is ahead by a small margin on the family mean and significantly on the set mean (by 0.011). It also costs about 20\% less, \$0.99 against \$1.25 per 1,000 queries for Perplexity v1.3 per passage. This lead comes from the German sets cut to 1,024 tokens, where every system scores below a good first stage. Averaged over the other five families, the two variants are level (0.540 each), although one question per passage is ahead on the full documents by 0.021 [0.004, 0.038]. Multiple choice is therefore the better default mainly because it costs less at the same quality. It also needs only one request per query, which keeps it clear of limits on requests per minute. Under a limit on tokens per minute it is the better choice as well, since one question per passage sends two to four times the tokens on the shortest passages. For a model priced per request instead of per token, such as Parallel's decisions-fast at \$0.001 per request~\cite{openrouter2026models}, one question per passage would cost $n$ times as much. None of the tested decision models is priced this way. Multiple choice has two conditions, however. The candidates should arrive in first-stage order, because some models lean on that order (Section~\ref{sec:position}), and the request must fit the model, or the query falls back to one question per passage. Without that order, the advantage of multiple choice is gone. With shuffled candidates, Perplexity v1.3 with multiple choice reaches a family mean of 0.520, below the 0.532 of one question per passage. One yes/no question per passage is free of both conditions. Its scores do not depend on the order of the candidates, and most requests stay small. It is therefore the safe choice when the candidate order cannot be trusted, for example when candidates are pooled from several sources without a merged ranking, or when the requests grow too large (Section~\ref{sec:size}). With Perplexity v1.2, one question per passage is ahead by a small margin on the family mean, while the set means are level. Perplexity v1.3 also leans much less on the candidate order than v1.2 (Section~\ref{sec:position}). Drex Max v1.1, the third model that ran in both variants on all sets, shows why each model needs a test of both variants. With multiple choice, it is 0.080 [0.073, 0.087] below its per-passage setup on the family mean, mostly on large requests (Section~\ref{sec:size}). Based on Perplexity v1.2 as well as v1.3, I therefore recommend multiple choice in first-stage order as the default, one yes/no question per passage as the fallback and a short test of both whenever a new model or model version appears.

\paragraph{At the same size.} Dinzinger et al.\ found that up to 9B, a dedicated reranker beats a decision model built on the same base model~\cite{dinzinger2026notbetter}. The data hold one comparison at this size, though with different base models. StartLux v1.1 9B is level with Qwen3 Reranker 8B on the family mean (0.524 against 0.522) and significantly behind Nemotron 3.5 Rerank 8B (0.536). On BRIGHT, however, it is significantly ahead of both (0.246 against 0.204 and 0.205).

\subsection{Uses beyond document reranking}

Reranking documents is one case of a more general question. Which of these items fits the current task? The same question appears in several places of an application built on language models, and a decision model can answer each of them with the two variants used in this paper.

\begin{itemize}
\item \textbf{Tool search.} An assistant with hundreds of tools should not load every tool definition up front, because the definitions fill its context while the model picks the right tool less reliably~\cite{anthropic2026toolsearch}. Vendors therefore offer a tool search tool, where the model asks for tools, then receives the best matches~\cite{anthropic2026toolsearch,openai2026toolsearch}. Open WebUI ships such a search as an experimental feature, with keyword search, embeddings or a decision model as its engine~\cite{openwebui2026release,openwebui2026toolsearch,openwebui2026toolsearchcommit}.
\item \textbf{Memory selection.} An application with long-term memory has to decide which stored memories to add to a conversation for the current message, and which to return when the model searches its memory.
\item \textbf{Memory review.} After a conversation, a background job can ask of each related memory whether it needs a change. This is a plain yes/no decision.
\item \textbf{Classification.} A decision model can sort texts into categories, and models that read images can do the same with images. It brings the broad world knowledge of a language model where a classic classifier needs its own training data, and it can run on one's own hardware.
\item \textbf{Context pruning.} In a long agent session, a decision model could mark which parts of the history no longer matter, such as failed tool calls or tool outputs that added nothing. These parts can then be dropped before or instead of a compaction of the context.
\end{itemize}

\medskip
An application could thus use one inexpensive model both for ranking and for such decisions.

None of these uses was evaluated here. The measurements cover document reranking, and they come with a warning for the list above. Tools and memories usually arrive in no meaningful order, which is exactly the case where some models lost quality with multiple choice (Section~\ref{sec:position}). For these uses, one yes/no question per item is the safer start.

\subsection{Outlook}

Decision models are a young category. The first one appeared on OpenRouter on 18 September 2026. On 5 November, the date of the listing the models were drawn from, OpenRouter listed at least 29 decision models. Only eight days later it listed 35~\cite{openrouter2026models}. In those eight days alone, Google, Perplexity, HeyGen, Parallel and Nace added seven models. One model, Mercury Decide, had left the listing by 9 November~\cite{openrouter2026mercury}. Vendors compete on price, speed and quality, releasing new versions within weeks. OpenRouter listed v1.3 of both Perplexity models 27 days after v1.2. Perplexity kept the sizes and list prices~\cite{openrouter2026models,perplexity2026changelog,perplexity2026pricing}. Nace and Liquid each listed three or more models within five weeks. I expect the reranking quality of decision models to rise at this pace, much as general language models improved over the past years, and prices to stay under pressure.

The dependence on the candidate order may also shrink. On the three shuffled sets, Gemini Decision 1 already shows no measurable dependence on the order. GPT-6 Luna depends on it only through its tied scores. From Perplexity v1.2 to v1.3, the loss from shuffling on SciFact fell by more than half (0.130 to 0.049). From Perplexity Flash v1.2 to Perplexity Flash v1.3, it did not fall (0.032 to 0.036, Section~\ref{sec:position}). If this continues, multiple choice will no longer need the candidates in first-stage order, which already holds for Gemini Decision 1.

These expectations go beyond the measurements. The data contain one version step, from v1.2 to v1.3, measured for both Perplexity models in the same setups. Perplexity Flash v1.3 improved clearly, while Perplexity v1.3 changed only by a small margin but leans much less on the candidate order (Section~\ref{sec:results}). A newer model can also score lower than an older one, depending on the variant it is asked in. Drex Max v1.1 came out after Drex v2.0, despite its lower version number. With multiple choice, it is 0.054 below Drex v2.0 with one question per passage on the family mean. The gap comes from BEIR, BRIGHT and the full documents, while on the other three families neither is significantly ahead. Asked one question per passage like Drex v2.0, Drex Max v1.1 is 0.026 [0.020, 0.031] ahead. A new version number guarantees nothing, so each new model, and each variant, needs its own test on retrieval data. A general index of decision quality is no substitute for that test (Section~\ref{sec:aa}).

\section{Limitations}
\label{sec:limitations}

\paragraph{Selection of models.} The decision models are a hand-picked selection from the OpenRouter listing of 5 November 2026, plus five later additions up to 13 November. Two of these, Drex Max v1.1 and StartLux v1.1 9B, are not on OpenRouter at all. OpenRouter lists Drex Max only in an earlier version. Three models were screened out by a pilot on a single set, while ten entered without screening, five from the listing plus the five later additions. Seven listed models that accept a free-form state were not run, among them Baseten mammoth-1, which stood high on the Artificial Analysis Decision Index (Section~\ref{sec:setup}). Gemini Decision 1, Perplexity v1.3 and Perplexity Flash v1.3 were added within a day of their release, a week after the listing date, while new models keep arriving every week (Section~\ref{sec:discussion}). Statements about ``decision models'' in this paper are therefore statements about the 19 tested models. The screening pilot used SciFact, and the choice of variant for some models used SciFact as well as the 100 German queries of BM25 sample 1. Both are part of the 35 sets. The German queries also occur in five other sets, so together with SciFact they carry about 10\% of the weight in the family mean and 12\% in the set mean.

\paragraph{One variant per model.} Perplexity v1.2 and v1.3 as well as Drex Max v1.1 ran in both variants. Eleven models were tested with one question per passage only, five with multiple choice only. Of these, the pilot compared both variants for Perplexity Flash v1.2 as well as GPT-6 Luna, and a check on SciFact did so for Gemini Decision 1. The pilot used 20 candidates per query. With 100 candidates, the two variants of Perplexity v1.2 differ only by a small margin, which fits the pilot. Drex Max v1.1 shows how much the variant can matter. It ran per passage only after its weak multiple-choice result, and it gained 0.080 on the family mean. Liquid d1.1 fell back to one question per passage for a fifth of its queries, so its row mixes both variants. The multiple-choice rows of the four Perplexity models also mix both on Robust04 with full documents, where 58 of the 249 queries fell back.

\paragraph{Single runs.} The main tables use one run of each system on each set. The significance test covers the choice of queries, but not changes the provider makes to a model between runs. Only 14 of the 28 models (19 decision models and nine dedicated rerankers) reported which exact version answered the requests (Table~\ref{tab:versions}). Perplexity v1.3 as well as Perplexity Flash v1.3 ran on a free launch offer, and it is unknown whether the free endpoint serves exactly the same models as the paid one. Nemotron 3.5 Rerank 8B likewise ran only on OpenRouter's free tier, which read at most 28,000 tokens of each document, while the tables price it at DeepInfra, whose endpoint was not used.

\paragraph{Many comparisons.} With so many comparisons, a few will look significant by chance alone. The wording rule of Section~\ref{sec:setup} keeps differences below 0.01 from being called significant, but a single interval that barely excludes zero should still be read with care. The set-mean lead of multiple choice for Perplexity v1.3 (0.011 [0.007, 0.015]) comes from the four German sets cut to 1,024 tokens. Without them, it is 0.004 and level. The bootstrap also resamples each set on its own, although 17 of the 35 sets reuse the queries of MLDR German, TREC-News, Robust04 or BRIGHT TheoremQA. MLDR German, TREC-News and Robust04 alone hold 706 queries, about a fifth of the distinct ones, yet they carry 54\% of the weight of the family mean (43\% of the set mean). The intervals of both means are therefore too narrow. Resampling the shared queries together leaves the set-mean lead of multiple choice for Perplexity v1.3 significant ($+0.011$ [$+0.006$, $+0.015$]), but other close intervals in this paper might include zero.

\paragraph{Order and ties.} Every number in the main tables breaks ties by the first-stage order, and the multiple-choice numbers also label the candidates in that order. Section~\ref{sec:position} shows that the first-stage order lifts Perplexity v1.3 by 0.018 on the family mean, mostly on the sets with 100 candidates outside BRIGHT. On SciFact, the order also lifts Perplexity v1.2 (by 0.130) and Drex Max v1.1 heavily (by 0.345). GPT-6 Luna gains from it through its many ties. Jev 1.14 rounds its scores to two decimals. Cohere Rerank 4 Pro, Cohere Rerank 4 Fast and Qwen3 Reranker 8B often give several candidates of a query the same score. On SciFact, the median query gets 57 different scores out of 100 from Cohere Rerank 4 Pro, 59 from Cohere Rerank 4 Fast and about 83 from Qwen3 Reranker 8B. Where measured, on 9 to 28 sets per model, breaking their ties randomly instead of by the first-stage order changes the three dedicated rerankers by at most 0.010 on a set, but lowers Jev 1.14 on Touch\'e-2020 from 0.346 to 0.279 (Section~\ref{sec:position}). The shuffle control covers all 35 sets for Perplexity v1.3 with multiple choice, three sets for five other multiple-choice models, only SciFact for Drex Max v1.1 and none for Liquid d1.1.

\paragraph{Benchmarks.} All benchmarks are public, so their documents or judgements may have been part of the training data of any system. The dedicated rerankers are usually trained on MS MARCO, the source of TREC DL. The training data of the decision models are not known, so the risk applies to them as well. Microsoft Decision 1, the best system on TREC DL, comes from the vendor whose search engine Bing supplied the queries of MS MARCO~\cite{bajaj2016msmarco}. The queries of MLDR were generated from paragraphs of the documents, which may favour lexical matching and puts the answer anywhere in a document. Where answers sit near the start, cutting documents short may hurt less. The chunk sets give each chunk the label of its document, so their results come out higher than they would if each chunk were judged on its own. Every system is scored with the same labels, but a system that finds the one chunk that answers the query earns no more than one that finds any chunk of a relevant document. On two BRIGHT sets, the first stage rarely finds a relevant document. The 35 sets cover English and German only.

\paragraph{Newer rerankers.} Cohere released Rerank 5 on 21 October 2026~\cite{cohere2026rerank5}. OpenRouter listed its Pro and Fast models on 3 November~\cite{openrouter2026rerank}, but they were not tested. Cohere states that Rerank 5 Fast outperforms Cohere Rerank 4 Pro~\cite{cohere2026rerank5}, so the best commercial rerankers of today may be ahead of those tested here.

\paragraph{Google's semantic ranker.} It ran on only 14 of the 35 sets, so it is missing from the overall table.

\paragraph{Cost and latency.} Costs are computed at list prices as of 13 November 2026, mostly through one routing service, without fees or tax. The same model can have a lower price elsewhere. DeepInfra charges a quarter of OpenRouter's price for Qwen3 Reranker 8B~\cite{deepinfra2026qwen}, while Nace lowered its own price for Drex v2.0 to half of OpenRouter's during November~\cite{nace2026pricing}. Systems that ran on a free offer or on my own hardware are shown at the public list price of the same model, which for Nemotron 3.5 Rerank 8B and bge-reranker-v2-m3 is the price of another provider. Free tiers and starting credits can change. Hardware and power for self-hosted systems are not priced. Latency was not measured under controlled conditions. For one question per passage, the latency with all requests sent at once is estimated from the slowest request of each query, because the main runs sent at most eight requests at a time (twelve for Perplexity v1.3).

\section{Conclusion}
\label{sec:conclusion}

A decision model becomes a reranker with a single question. On 35 test sets, the best decision models, two Perplexity setups, are as good as the best commercial rerankers tested, at 20 to 45\% of their cost. Decision models lead on the reasoning-heavy BRIGHT benchmark, while dedicated rerankers keep a small lead on BEIR as well as on long German documents read to 8,192 tokens. On long German documents, how much a reranker reads matters more than which good reranker it is. The method needs neither training nor prompt tuning. Both prompts were written once. For Perplexity v1.2 and Perplexity Flash v1.2, five other prompts for the multiple-choice question changed the result by about 0.01 or less. For Perplexity v1.2 and v1.3, multiple choice with the candidates in first-stage order is a good default because it costs less at about the same quality. Perplexity v1.3 favours it on the set mean, while with v1.2 both variants score almost the same.

The category is still uneven. Some decision models rerank poorly, two refuse queries or have them blocked and one falls below the first stage on many sets with 100 candidates when asked with multiple choice. The top decision model of the Artificial Analysis Decision Index is the weakest reranker among the indexed models tested when asked with multiple choice. With the candidates in random order, it then loses two thirds of its result on SciFact. Asked one question per passage, which costs more, it is third of the 14. A high position on such an index thus gives no assurance for reranking. The category also changes within weeks, as the gain of 0.022 from Perplexity Flash v1.2 to Perplexity Flash v1.3 shows. Anyone who wants to use a decision model as a reranker should therefore test it on retrieval data, with the variant and the request size of the intended use. Once that test is done, the best decision models are a cheap alternative to dedicated rerankers.

\section*{Acknowledgements}
I thank Harald Lampesberger (University of Applied Sciences Upper Austria, Hagenberg) for feedback on this work, for API credits and for an API key for Nemotron 3.5 Rerank 8B. I thank Pkeffect, Kemon and Mateo from LiteLLM for further free API keys for Nemotron 3.5 Rerank 8B.

\section*{Funding and costs}
This study received no funding. The API usage for this paper actually cost \$441 as charged by the providers, according to the per-request records as well as the account statements, including pilots, controls and repeated runs. Harald Lampesberger contributed \$18.98 of it in API credits, and I paid the rest. In total, I paid \texteuro{}520.31 for API credits, fees and taxes, more than the \$441 of usage. A small part of the credit is still unused. Some runs used free offers, namely the free tier of Nemotron 3.5 Rerank 8B on OpenRouter, the free starting credit of Nace, the free tier of Google AI Studio, as well as the free launch offer of Perplexity v1.3 and Perplexity Flash v1.3 on OpenRouter. At list prices, these runs and the self-hosted bge-reranker-v2-m3 would have cost a further \$70, of which \$59.88 fall on the single pass of all systems in Table~\ref{tab:cost}. Paid and free runs together thus come to about \$510. The self-hosted models ran on my own consumer hardware. Also not included are the 2,706 queries to Google's semantic ranker (\$2.71 at its list price), which ran through a proxy at no cost to me, and the mistral-embed embeddings, which ran on a subscription.

\section*{Use of AI tools}
I used Claude Code with Claude Opus 5.5 to write and run the experiment scripts. It also helped me write the text and check it against the data as well as the cited sources.

\bibliographystyle{plainnat}
\bibliography{refs}

\clearpage
\appendix
\section{Results per set}
\label{app:perset}

Tables~\ref{tab:perset-web}, \ref{tab:perset-bright} and~\ref{tab:perset-long} give \ndcg{} for every system on each of the 35 sets, in the order of Table~\ref{tab:overview}.

\begin{table}[!htbp]
\centering
\scriptsize
\setlength{\tabcolsep}{2.4pt}
\begin{tabular}{lcccccccccc}
\toprule
System & DL19 & DL20 & Covid & NFC & Touch\'e & DBP & SciF & Signal & News & R04 \\
\midrule
First stage only & 0.506 & 0.480 & 0.595 & 0.337 & 0.442 & 0.318 & 0.679 & 0.331 & 0.395 & 0.407 \\
\midrule
Voyage Rerank 3 & 0.759 & 0.725 & 0.861 & 0.400 & 0.329 & 0.479 & 0.791 & 0.324 & 0.521 & 0.639 \\
Perplexity v1.2 (A) & 0.733 & 0.714 & 0.834 & 0.399 & 0.282 & 0.459 & 0.800 & 0.294 & 0.503 & 0.654 \\
Cohere Rerank 4 Pro & 0.749 & 0.709 & 0.852 & 0.402 & 0.391 & 0.467 & 0.800 & 0.320 & 0.526 & 0.619 \\
Perplexity v1.3 (C) & 0.721 & 0.703 & 0.797 & 0.390 & 0.354 & 0.456 & 0.802 & 0.341 & 0.485 & 0.609 \\
Voyage Rerank 2.5 & 0.742 & 0.718 & \textbf{0.868} & 0.405 & 0.332 & 0.467 & 0.785 & 0.325 & 0.505 & 0.635 \\
Nemotron 3.5 Rerank 8B & 0.730 & 0.706 & 0.857 & \textbf{0.411} & 0.343 & \textbf{0.490} & \textbf{0.820} & 0.342 & \textbf{0.528} & 0.628 \\
Microsoft Decision 1 (A) & \textbf{0.771} & \textbf{0.761} & 0.815 & 0.392 & 0.303 & 0.453 & 0.784 & 0.290 & 0.498 & 0.641 \\
Drex Max v1.1 (A) & 0.726 & 0.705 & 0.851 & 0.402 & 0.263 & 0.456 & 0.793 & 0.314 & 0.499 & 0.648 \\
Liquid d1-ultra (A) & 0.736 & 0.697 & 0.831 & 0.403 & 0.293 & 0.457 & 0.788 & 0.303 & 0.497 & \textbf{0.659} \\
Perplexity v1.2 (C) & 0.707 & 0.695 & 0.789 & 0.379 & 0.376 & 0.458 & 0.787 & \textbf{0.345} & 0.475 & 0.584 \\
Perplexity v1.3 (A) & 0.722 & 0.703 & 0.833 & 0.399 & 0.266 & 0.452 & 0.799 & 0.290 & 0.509 & 0.657 \\
Liquid d1 (A) & 0.739 & 0.710 & 0.832 & 0.396 & 0.316 & 0.442 & 0.788 & 0.302 & 0.494 & 0.641 \\
Jev 1.14 (A) & 0.728 & 0.691 & 0.830 & 0.395 & 0.346 & 0.415 & 0.801 & 0.292 & 0.478 & 0.645 \\
Morph Reflex (A) & 0.728 & 0.694 & 0.843 & 0.398 & 0.309 & 0.457 & 0.792 & 0.287 & 0.474 & 0.632 \\
elephant-1 (A) & 0.726 & 0.706 & 0.824 & 0.397 & 0.327 & 0.434 & 0.778 & 0.314 & 0.483 & 0.636 \\
StartLux v1.1 9B (A) & 0.728 & 0.687 & 0.814 & 0.395 & 0.312 & 0.442 & 0.796 & 0.321 & 0.497 & 0.621 \\
Cohere Rerank 4 Fast & 0.748 & 0.700 & 0.841 & 0.389 & 0.363 & 0.453 & 0.775 & 0.318 & 0.507 & 0.594 \\
Qwen3 Reranker 8B & 0.732 & 0.672 & 0.846 & 0.393 & \textbf{0.411} & 0.460 & 0.798 & 0.331 & 0.509 & 0.587 \\
Mercury Decide (A) & 0.711 & 0.694 & 0.822 & 0.393 & 0.275 & 0.442 & 0.784 & 0.305 & 0.473 & 0.643 \\
Liquid d1.1 (C) & 0.708 & 0.686 & 0.795 & 0.392 & 0.339 & 0.433 & 0.804 & 0.310 & 0.457 & 0.604 \\
Drex v2.0 (A) & 0.709 & 0.682 & 0.816 & 0.382 & 0.306 & 0.395 & 0.773 & 0.258 & 0.435 & 0.623 \\
Perplexity Flash v1.3 (C) & 0.716 & 0.691 & 0.821 & 0.394 & 0.325 & 0.448 & 0.795 & 0.318 & 0.455 & 0.595 \\
GPT-6 Luna (C) & 0.651 & 0.619 & 0.762 & 0.376 & 0.353 & 0.414 & 0.787 & 0.320 & 0.425 & 0.552 \\
Gemini Decision 1 (C) & 0.711 & 0.711 & 0.812 & 0.393 & 0.244 & 0.463 & 0.800 & 0.340 & 0.478 & 0.596 \\
Cohere Rerank 3.5 & 0.728 & 0.683 & 0.799 & 0.367 & 0.364 & 0.452 & 0.765 & 0.309 & 0.489 & 0.555 \\
Perplexity Flash v1.2 (C) & 0.695 & 0.684 & 0.825 & 0.382 & 0.323 & 0.453 & 0.780 & 0.326 & 0.445 & 0.551 \\
Solar Decide Flash (A) & 0.669 & 0.639 & 0.797 & 0.380 & 0.248 & 0.386 & 0.720 & 0.272 & 0.445 & 0.615 \\
bge-reranker-v2-m3 & 0.724 & 0.681 & 0.779 & 0.356 & 0.357 & 0.444 & 0.734 & 0.291 & 0.460 & 0.521 \\
Drex Max v1.1 (C) & 0.693 & 0.692 & 0.663 & 0.260 & 0.227 & 0.462 & 0.526 & 0.344 & 0.279 & 0.315 \\
joey-1 (A) & 0.701 & 0.660 & 0.805 & 0.362 & 0.266 & 0.398 & 0.676 & 0.259 & 0.390 & 0.555 \\
MiniLM-L6 & 0.727 & 0.675 & 0.745 & 0.368 & 0.276 & 0.442 & 0.687 & 0.339 & 0.434 & 0.473 \\
\bottomrule
\end{tabular}
\caption{\ndcg{} per set, TREC DL and BEIR (BM25 top 100, passages cut to 512 tokens). Sorted by the family mean, as in Table~\ref{tab:overview}. Bold: best system of the column, without the first-stage row. Columns: TREC DL 2019, TREC DL 2020, TREC-COVID, NFCorpus, Touch\'e-2020, DBPedia, SciFact, Signal-1M, TREC-News, Robust04.}
\label{tab:perset-web}
\end{table}

\begin{table}[!htbp]
\centering
\scriptsize
\setlength{\tabcolsep}{2.4pt}
\begin{tabular}{lcccccccccccc}
\toprule
System & Bio & Earth & Econ & Psy & Rob & Stack & Sust & Leet & Pony & AoPS & ThQ & ThT \\
\midrule
First stage only & 0.181 & 0.281 & 0.164 & 0.134 & 0.109 & 0.162 & 0.161 & 0.247 & 0.043 & 0.065 & 0.073 & 0.021 \\
\midrule
Voyage Rerank 3 & 0.295 & 0.406 & 0.241 & 0.297 & 0.240 & 0.233 & 0.246 & 0.303 & 0.181 & 0.123 & 0.122 & 0.110 \\
Perplexity v1.2 (A) & 0.339 & 0.425 & 0.266 & 0.325 & 0.268 & 0.248 & 0.302 & 0.322 & 0.201 & 0.107 & 0.117 & 0.117 \\
Cohere Rerank 4 Pro & 0.312 & 0.435 & 0.240 & 0.307 & 0.242 & 0.233 & 0.267 & 0.275 & 0.134 & 0.122 & 0.133 & 0.119 \\
Perplexity v1.3 (C) & 0.336 & 0.440 & \textbf{0.287} & 0.331 & 0.288 & 0.263 & 0.309 & \textbf{0.376} & 0.115 & \textbf{0.148} & \textbf{0.142} & 0.119 \\
Voyage Rerank 2.5 & 0.302 & 0.418 & 0.245 & 0.293 & 0.250 & 0.234 & 0.269 & 0.277 & 0.200 & 0.108 & 0.118 & 0.107 \\
Nemotron 3.5 Rerank 8B & 0.218 & 0.381 & 0.208 & 0.261 & 0.239 & 0.200 & 0.212 & 0.286 & 0.107 & 0.125 & 0.117 & 0.102 \\
Microsoft Decision 1 (A) & 0.325 & 0.409 & 0.253 & 0.307 & 0.259 & 0.239 & 0.298 & 0.340 & 0.179 & 0.119 & 0.120 & 0.109 \\
Drex Max v1.1 (A) & \textbf{0.341} & 0.409 & 0.249 & 0.317 & 0.250 & 0.232 & 0.288 & 0.278 & \textbf{0.214} & 0.097 & 0.119 & 0.112 \\
Liquid d1-ultra (A) & 0.323 & 0.403 & 0.286 & 0.325 & 0.265 & 0.244 & 0.292 & 0.283 & 0.196 & 0.112 & 0.116 & 0.113 \\
Perplexity v1.2 (C) & 0.332 & 0.445 & 0.270 & \textbf{0.333} & 0.278 & 0.255 & 0.303 & 0.362 & 0.111 & 0.140 & 0.138 & 0.116 \\
Perplexity v1.3 (A) & 0.331 & 0.440 & 0.276 & 0.320 & 0.269 & 0.256 & 0.314 & 0.335 & 0.206 & 0.104 & 0.128 & 0.121 \\
Liquid d1 (A) & 0.331 & 0.461 & 0.273 & 0.327 & 0.267 & 0.251 & 0.279 & 0.293 & 0.141 & 0.097 & 0.111 & 0.112 \\
Jev 1.14 (A) & 0.319 & 0.426 & 0.279 & 0.299 & 0.253 & 0.250 & 0.300 & 0.334 & 0.192 & 0.111 & 0.118 & 0.122 \\
Morph Reflex (A) & 0.312 & 0.424 & 0.252 & 0.318 & 0.238 & 0.259 & 0.295 & 0.299 & 0.204 & 0.090 & 0.115 & 0.115 \\
elephant-1 (A) & 0.296 & 0.378 & 0.242 & 0.295 & 0.236 & 0.243 & 0.266 & 0.332 & 0.203 & 0.114 & 0.118 & 0.106 \\
StartLux v1.1 9B (A) & 0.336 & 0.445 & 0.264 & 0.311 & 0.258 & 0.245 & 0.287 & 0.298 & 0.179 & 0.085 & 0.126 & 0.116 \\
Cohere Rerank 4 Fast & 0.310 & 0.402 & 0.238 & 0.277 & 0.203 & 0.221 & 0.257 & 0.269 & 0.149 & 0.129 & 0.126 & \textbf{0.124} \\
Qwen3 Reranker 8B & 0.227 & 0.355 & 0.229 & 0.260 & 0.250 & 0.200 & 0.225 & 0.288 & 0.099 & 0.108 & 0.111 & 0.094 \\
Mercury Decide (A) & 0.296 & 0.418 & 0.275 & 0.315 & 0.232 & 0.236 & 0.284 & 0.318 & 0.173 & 0.108 & 0.117 & 0.121 \\
Liquid d1.1 (C) & 0.332 & 0.404 & \textbf{0.287} & 0.322 & 0.258 & 0.234 & 0.314 & 0.308 & 0.117 & 0.134 & 0.134 & 0.118 \\
Drex v2.0 (A) & 0.337 & 0.445 & 0.265 & 0.308 & 0.261 & 0.256 & 0.298 & 0.278 & 0.182 & 0.081 & 0.106 & 0.101 \\
Perplexity Flash v1.3 (C) & 0.328 & 0.361 & 0.272 & 0.294 & 0.253 & 0.229 & 0.286 & 0.265 & 0.111 & 0.121 & 0.121 & 0.119 \\
GPT-6 Luna (C) & 0.315 & 0.444 & \textbf{0.287} & 0.314 & \textbf{0.291} & \textbf{0.275} & 0.300 & 0.267 & 0.139 & 0.124 & 0.132 & 0.116 \\
Gemini Decision 1 (C) & 0.340 & \textbf{0.465} & 0.282 & 0.324 & 0.272 & 0.228 & \textbf{0.324} & 0.294 & 0.196 & 0.131 & 0.131 & 0.105 \\
Cohere Rerank 3.5 & 0.183 & 0.284 & 0.145 & 0.206 & 0.144 & 0.150 & 0.140 & 0.246 & 0.052 & 0.085 & 0.087 & 0.078 \\
Perplexity Flash v1.2 (C) & 0.313 & 0.319 & 0.223 & 0.273 & 0.197 & 0.187 & 0.255 & 0.196 & 0.085 & 0.088 & 0.100 & 0.115 \\
Solar Decide Flash (A) & 0.258 & 0.336 & 0.164 & 0.237 & 0.095 & 0.138 & 0.157 & 0.230 & 0.089 & 0.071 & 0.090 & 0.104 \\
bge-reranker-v2-m3 & 0.164 & 0.337 & 0.082 & 0.140 & 0.113 & 0.118 & 0.144 & 0.171 & 0.066 & 0.065 & 0.076 & 0.045 \\
Drex Max v1.1 (C) & 0.234 & 0.234 & 0.140 & 0.173 & 0.122 & 0.090 & 0.234 & 0.175 & 0.087 & 0.080 & 0.059 & 0.020 \\
joey-1 (A) & 0.223 & 0.309 & 0.153 & 0.204 & 0.129 & 0.150 & 0.198 & 0.227 & 0.094 & 0.088 & 0.087 & 0.112 \\
MiniLM-L6 & 0.118 & 0.188 & 0.123 & 0.118 & 0.150 & 0.123 & 0.116 & 0.221 & 0.123 & 0.089 & 0.053 & 0.028 \\
\bottomrule
\end{tabular}
\caption{\ndcg{} per set, BRIGHT (BM25 top 100, passages cut to 512 tokens). Sorted by the family mean, as in Table~\ref{tab:overview}. Bold: best system of the column, without the first-stage row. Columns: Biology, Earth Science, Economics, Psychology, Robotics, Stack Overflow, Sustainable Living, LeetCode, Pony, AoPS, TheoremQA-Questions, TheoremQA-Theorems.}
\label{tab:perset-bright}
\end{table}

\begin{table}[!htbp]
\centering
\scriptsize
\setlength{\tabcolsep}{2.4pt}
\resizebox{\textwidth}{!}{%
\begin{tabular}{lccccccccccccc}
\toprule
System & B1/1k & B1/8k & B2/1k & B2/8k & M/1k & M/8k & Q/1k & Q/8k & C-MLDR & C-News & C-R04 & News-F & R04-F \\
\midrule
First stage only & 0.564 & 0.564 & 0.655 & 0.655 & 0.257 & 0.257 & 0.550 & 0.550 & 0.401 & 0.658 & 0.677 & 0.395 & 0.407 \\
\midrule
Voyage Rerank 3 & 0.430 & 0.663 & 0.483 & 0.697 & 0.313 & 0.360 & 0.455 & 0.656 & 0.451 & 0.673 & 0.761 & 0.518 & 0.659 \\
Perplexity v1.2 (A) & 0.475 & 0.640 & 0.523 & 0.695 & 0.309 & 0.357 & 0.465 & 0.646 & 0.443 & 0.661 & \textbf{0.774} & 0.519 & 0.676 \\
Cohere Rerank 4 Pro & 0.469 & 0.639 & 0.526 & 0.677 & 0.316 & 0.354 & 0.479 & 0.639 & 0.447 & \textbf{0.678} & 0.742 & \textbf{0.535} & 0.649 \\
Perplexity v1.3 (C) & 0.513 & 0.655 & 0.558 & 0.686 & 0.321 & 0.356 & 0.505 & 0.650 & 0.468 & 0.662 & 0.756 & 0.492 & 0.643 \\
Voyage Rerank 2.5 & 0.441 & \textbf{0.670} & 0.510 & 0.712 & 0.307 & \textbf{0.364} & 0.454 & \textbf{0.664} & 0.448 & 0.658 & 0.757 & 0.520 & 0.655 \\
Nemotron 3.5 Rerank 8B & 0.495 & 0.650 & 0.553 & 0.686 & 0.321 & 0.357 & 0.488 & 0.638 & 0.458 & 0.673 & 0.750 & 0.532 & 0.650 \\
Microsoft Decision 1 (A) & 0.442 & 0.629 & 0.461 & 0.672 & 0.286 & 0.354 & 0.416 & 0.633 & 0.429 & 0.647 & 0.768 & 0.488 & 0.663 \\
Drex Max v1.1 (A) & 0.471 & 0.641 & 0.512 & 0.698 & 0.314 & 0.358 & 0.471 & 0.650 & 0.443 & 0.654 & 0.767 & 0.498 & 0.672 \\
Liquid d1-ultra (A) & 0.467 & 0.628 & 0.514 & 0.697 & 0.307 & 0.357 & 0.457 & 0.649 & 0.445 & 0.646 & 0.770 & 0.485 & \textbf{0.681} \\
Perplexity v1.2 (C) & \textbf{0.532} & 0.642 & 0.580 & 0.693 & \textbf{0.328} & 0.357 & \textbf{0.524} & 0.653 & \textbf{0.472} & 0.651 & 0.750 & 0.497 & 0.620 \\
Perplexity v1.3 (A) & 0.432 & 0.626 & 0.470 & 0.686 & 0.299 & 0.357 & 0.431 & 0.638 & 0.436 & 0.645 & 0.772 & 0.497 & 0.680 \\
Liquid d1 (A) & 0.442 & 0.622 & 0.487 & 0.676 & 0.304 & 0.348 & 0.443 & 0.613 & 0.440 & 0.644 & 0.757 & 0.486 & 0.662 \\
Jev 1.14 (A) & 0.438 & 0.625 & 0.518 & 0.673 & 0.300 & 0.347 & 0.427 & 0.628 & 0.429 & 0.649 & 0.769 & 0.475 & 0.666 \\
Morph Reflex (A) & 0.433 & 0.629 & 0.490 & 0.687 & 0.297 & 0.356 & 0.432 & 0.638 & 0.428 & 0.633 & 0.761 & 0.476 & 0.662 \\
elephant-1 (A) & 0.404 & 0.626 & 0.453 & 0.669 & 0.280 & 0.346 & 0.413 & 0.625 & 0.422 & 0.648 & 0.754 & 0.499 & 0.666 \\
StartLux v1.1 9B (A) & 0.428 & 0.613 & 0.460 & 0.648 & 0.303 & 0.348 & 0.433 & 0.610 & 0.428 & 0.662 & 0.744 & 0.501 & 0.643 \\
Cohere Rerank 4 Fast & 0.432 & 0.630 & 0.504 & 0.656 & 0.316 & 0.350 & 0.444 & 0.612 & 0.439 & 0.637 & 0.722 & 0.507 & 0.619 \\
Qwen3 Reranker 8B & 0.449 & 0.664 & 0.541 & \textbf{0.721} & 0.324 & 0.361 & 0.470 & 0.662 & 0.448 & 0.636 & 0.719 & 0.516 & 0.597 \\
Mercury Decide (A) & 0.441 & 0.627 & 0.503 & 0.684 & 0.290 & 0.353 & 0.428 & 0.623 & 0.430 & 0.612 & 0.762 & 0.466 & 0.664 \\
Liquid d1.1 (C) & 0.460 & 0.634 & 0.530 & 0.672 & 0.315 & 0.346 & 0.464 & 0.626 & 0.448 & 0.625 & 0.734 & 0.431 & 0.658 \\
Drex v2.0 (A) & 0.400 & 0.626 & 0.470 & 0.664 & 0.291 & 0.353 & 0.425 & 0.620 & 0.435 & 0.600 & 0.757 & 0.423 & 0.652 \\
Perplexity Flash v1.3 (C) & 0.438 & 0.503 & 0.478 & 0.568 & 0.305 & 0.345 & 0.437 & 0.559 & 0.441 & 0.623 & 0.752 & 0.437 & 0.562 \\
GPT-6 Luna (C) & 0.496 & 0.644 & 0.584 & 0.682 & 0.311 & 0.353 & 0.504 & 0.641 & 0.445 & 0.610 & 0.704 & 0.427 & 0.541 \\
Gemini Decision 1 (C) & 0.427 & 0.373 & 0.524 & 0.564 & 0.291 & 0.266 & 0.471 & 0.524 & 0.430 & 0.602 & 0.724 & 0.418 & 0.542 \\
Cohere Rerank 3.5 & 0.438 & 0.476 & 0.485 & 0.508 & 0.302 & 0.321 & 0.450 & 0.472 & 0.438 & 0.631 & 0.682 & 0.485 & 0.526 \\
Perplexity Flash v1.2 (C) & 0.441 & 0.466 & 0.497 & 0.523 & 0.300 & 0.334 & 0.447 & 0.505 & 0.430 & 0.625 & 0.721 & 0.410 & 0.490 \\
Solar Decide Flash (A) & 0.379 & 0.577 & 0.438 & 0.622 & 0.254 & 0.329 & 0.373 & 0.576 & 0.393 & 0.587 & 0.751 & 0.429 & 0.638 \\
bge-reranker-v2-m3 & 0.389 & 0.584 & 0.455 & 0.664 & 0.311 & 0.342 & 0.447 & 0.592 & 0.419 & 0.604 & 0.638 & 0.456 & 0.494 \\
Drex Max v1.1 (C) & 0.515 & 0.561 & \textbf{0.585} & 0.600 & 0.318 & 0.326 & 0.509 & 0.539 & 0.445 & 0.658 & 0.726 & 0.378 & 0.441 \\
joey-1 (A) & 0.364 & 0.469 & 0.422 & 0.557 & 0.247 & 0.302 & 0.367 & 0.506 & 0.368 & 0.557 & 0.663 & 0.359 & 0.542 \\
MiniLM-L6 & 0.283 & 0.283 & 0.308 & 0.308 & 0.223 & 0.223 & 0.342 & 0.342 & 0.324 & 0.577 & 0.591 & 0.434 & 0.473 \\
\bottomrule
\end{tabular}}
\caption{\ndcg{} per set, long German documents (B1, B2: the two BM25 samples, M: mistral-embed, Q: Qwen3 Embedding 8B, 1k: documents cut to 1,024 tokens, 8k: documents cut to 8,192 tokens), 2,000-token chunks (C-) and full documents (-F). Sorted by the family mean, as in Table~\ref{tab:overview}. Bold: best system of the column, without the first-stage row.}
\label{tab:perset-long}
\end{table}

\section{Cost per set}
\label{app:cost}

Table~\ref{tab:cost-sets} lists the cost of each set for one pass of all 31 systems.

\begin{table}[!htbp]
\centering
\footnotesize
\setlength{\tabcolsep}{3.5pt}
\begin{tabular}{llrrr}
\toprule
Family & Set & Queries & 31 systems & Google \\
\midrule
DL & TREC DL 2019 & 43 & 1.12 & 0.04 \\
DL & TREC DL 2020 & 54 & 1.43 & 0.05 \\
\midrule
BEIR & TREC-COVID & 50 & 3.02 & 0.05 \\
BEIR & NFCorpus & 308 & 14.20 & 0.31 \\
BEIR & Touch\'e-2020 & 49 & 3.37 & -- \\
BEIR & DBPedia & 400 & 10.61 & -- \\
BEIR & SciFact & 300 & 18.95 & 0.30 \\
BEIR & Signal-1M & 97 & 1.85 & -- \\
BEIR & TREC-News & 57 & 4.66 & -- \\
BEIR & Robust04 & 249 & 20.30 & -- \\
\midrule
BRIGHT & Biology & 103 & 4.59 & 0.10 \\
BRIGHT & Earth Science & 116 & 6.39 & -- \\
BRIGHT & Economics & 103 & 8.25 & -- \\
BRIGHT & Psychology & 101 & 6.73 & -- \\
BRIGHT & Robotics & 101 & 10.98 & -- \\
BRIGHT & Stack Overflow & 117 & 13.20 & -- \\
BRIGHT & Sustainable Living & 108 & 6.29 & -- \\
BRIGHT & LeetCode & 142 & 13.41 & 0.14 \\
BRIGHT & Pony & 112 & 5.38 & -- \\
BRIGHT & AoPS & 111 & 6.40 & -- \\
BRIGHT & TheoremQA-Questions & 194 & 11.65 & -- \\
BRIGHT & TheoremQA-Theorems & 76 & 5.43 & -- \\
\midrule
German & BM25 sample 1, 8,192 tokens & 100 & 21.84 & 0.10 \\
German & BM25 sample 1, 1,024 tokens & 100 & 3.43 & -- \\
German & BM25 sample 2, 8,192 tokens & 100 & 22.23 & 0.10 \\
German & BM25 sample 2, 1,024 tokens & 100 & 3.45 & -- \\
German & mistral-embed, 8,192 tokens & 400 & 38.76 & 0.40 \\
German & mistral-embed, 1,024 tokens & 400 & 9.95 & -- \\
German & Qwen3 Embedding 8B, 8,192 tokens & 400 & 85.89 & 0.40 \\
German & Qwen3 Embedding 8B, 1,024 tokens & 400 & 13.73 & -- \\
\midrule
Chunks & MLDR German chunks & 400 & 23.69 & 0.40 \\
Chunks & TREC-News chunks & 57 & 2.00 & 0.06 \\
Chunks & Robust04 chunks & 249 & 8.87 & 0.25 \\
\midrule
Full & TREC-News, full documents & 57 & 11.69 & -- \\
Full & Robust04, full documents & 249 & 64.87 & -- \\
\midrule
All & 35 sets & 6,003 & 488.60 & 2.71 \\
\bottomrule
\end{tabular}
\caption{Cost of each set at API list prices in US dollars, summed over the 31 systems for one pass, and Google's semantic ranker at its list price on the sets it ran. Totals are computed before rounding. Grouped by family in the column order of Table~\ref{tab:overview}.}
\label{tab:cost-sets}
\end{table}

\section{Shuffle control on every set}
\label{app:shuffle}

Table~\ref{tab:shuffle-sets} gives the shuffle control of Perplexity v1.3 with multiple choice for each of the 35 sets (Section~\ref{sec:position}).

\begin{table}[!htb]
\centering
\scriptsize
\setlength{\tabcolsep}{3.5pt}
\begin{tabular}{llcccl}
\toprule
Family & Set & First stage only & First-stage order & Shuffled & Change \\
\midrule
DL & TREC DL 2019 & 0.506 & 0.721 & 0.691 & $-$0.030 [$-$0.051, $-$0.010] \\
DL & TREC DL 2020 & 0.480 & 0.703 & 0.681 & $-$0.022 [$-$0.038, $-$0.005] \\
\midrule
BEIR & TREC-COVID & 0.595 & 0.797 & 0.771 & $-$0.026 [$-$0.054, $+$0.002] \\
BEIR & NFCorpus & 0.337 & 0.390 & 0.349 & $-$0.042 [$-$0.056, $-$0.028] \\
BEIR & Touch\'e-2020 & 0.442 & 0.354 & 0.329 & $-$0.024 [$-$0.048, $-$0.002] \\
BEIR & DBPedia & 0.318 & 0.456 & 0.438 & $-$0.019 [$-$0.027, $-$0.011] \\
BEIR & SciFact & 0.679 & 0.802 & 0.753 & $-$0.049 [$-$0.070, $-$0.028] \\
BEIR & Signal-1M & 0.331 & 0.341 & 0.307 & $-$0.035 [$-$0.054, $-$0.015] \\
BEIR & TREC-News & 0.395 & 0.485 & 0.440 & $-$0.045 [$-$0.067, $-$0.026] \\
BEIR & Robust04 & 0.407 & 0.609 & 0.573 & $-$0.036 [$-$0.048, $-$0.024] \\
\midrule
BRIGHT & Biology & 0.181 & 0.336 & 0.334 & $-$0.002 [$-$0.013, $+$0.009] \\
BRIGHT & Earth Science & 0.281 & 0.440 & 0.426 & $-$0.015 [$-$0.034, $+$0.003] \\
BRIGHT & Economics & 0.164 & 0.287 & 0.288 & $+$0.001 [$-$0.013, $+$0.016] \\
BRIGHT & Psychology & 0.134 & 0.331 & 0.321 & $-$0.010 [$-$0.027, $+$0.005] \\
BRIGHT & Robotics & 0.109 & 0.288 & 0.274 & $-$0.014 [$-$0.029, $-$0.002] \\
BRIGHT & Stack Overflow & 0.162 & 0.263 & 0.239 & $-$0.024 [$-$0.048, $-$0.005] \\
BRIGHT & Sustainable Living & 0.161 & 0.309 & 0.298 & $-$0.011 [$-$0.024, $-$0.000] \\
BRIGHT & LeetCode & 0.247 & 0.376 & 0.378 & $+$0.002 [$-$0.011, $+$0.016] \\
BRIGHT & Pony & 0.043 & 0.115 & 0.123 & $+$0.008 [$-$0.004, $+$0.019] \\
BRIGHT & AoPS & 0.065 & 0.148 & 0.149 & $+$0.001 [$-$0.007, $+$0.009] \\
BRIGHT & TheoremQA-Questions & 0.073 & 0.142 & 0.139 & $-$0.002 [$-$0.005, $+$0.000] \\
BRIGHT & TheoremQA-Theorems & 0.021 & 0.119 & 0.119 & $+$0.000 [$+$0.000, $+$0.000] \\
\midrule
German & BM25 sample 1, 8,192 tokens & 0.564 & 0.655 & 0.638 & $-$0.017 [$-$0.041, $+$0.008] \\
German & BM25 sample 1, 1,024 tokens & 0.564 & 0.513 & 0.490 & $-$0.022 [$-$0.044, $-$0.004] \\
German & BM25 sample 2, 8,192 tokens & 0.655 & 0.686 & 0.688 & $+$0.002 [$-$0.013, $+$0.017] \\
German & BM25 sample 2, 1,024 tokens & 0.655 & 0.558 & 0.540 & $-$0.019 [$-$0.045, $+$0.008] \\
German & mistral-embed, 8,192 tokens & 0.257 & 0.356 & 0.357 & $+$0.000 [$-$0.003, $+$0.004] \\
German & mistral-embed, 1,024 tokens & 0.257 & 0.321 & 0.323 & $+$0.002 [$-$0.005, $+$0.008] \\
German & Qwen3 Embedding 8B, 8,192 tokens & 0.550 & 0.650 & 0.654 & $+$0.004 [$-$0.003, $+$0.012] \\
German & Qwen3 Embedding 8B, 1,024 tokens & 0.550 & 0.505 & 0.485 & $-$0.020 [$-$0.032, $-$0.008] \\
\midrule
Chunks & MLDR German chunks & 0.401 & 0.468 & 0.463 & $-$0.005 [$-$0.010, $+$0.001] \\
Chunks & TREC-News chunks & 0.658 & 0.662 & 0.650 & $-$0.011 [$-$0.033, $+$0.007] \\
Chunks & Robust04 chunks & 0.677 & 0.756 & 0.751 & $-$0.005 [$-$0.013, $+$0.003] \\
\midrule
Full & TREC-News, full documents & 0.395 & 0.492 & 0.465 & $-$0.027 [$-$0.050, $-$0.004] \\
Full & Robust04, full documents & 0.407 & 0.643 & 0.612 & $-$0.031 [$-$0.042, $-$0.020] \\
\bottomrule
\end{tabular}
\caption{Perplexity v1.3 (C) on every set, once with the candidates labelled in first-stage order and once in a random order for each query, with \ndcg{} and the change with its 95\% interval (paired bootstrap). Ties are broken randomly on both sides. Queries whose request was too large ran with one question per passage in both runs, which does not depend on the order. Grouped by family in the column order of Table~\ref{tab:overview}.}
\label{tab:shuffle-sets}
\end{table}

\section{Prompt test}
\label{app:prompt}

Table~\ref{tab:prompt} shows the six prompts for the multiple-choice question and their results. The seven sets are TREC DL 2019, TREC DL 2020, SciFact, TREC-News and three BRIGHT sets (Biology, Economics, LeetCode).

\begin{table}[!htbp]
\centering
\footnotesize
\setlength{\tabcolsep}{3.5pt}
\resizebox{\textwidth}{!}{%
\begin{tabular}{lclcl}
\toprule
 & \multicolumn{2}{c}{Perplexity v1.2 (C)} & \multicolumn{2}{c}{Perplexity Flash v1.2 (C)} \\
\cmidrule(lr){2-3}\cmidrule(lr){4-5}
Question & \ndcg{} & vs.\ default & \ndcg{} & vs.\ default \\
\midrule
\parbox[t]{5.6cm}{\raggedright Which passage is the most relevant to the query?\strut} & 0.491 & -- & 0.452 & -- \\
\parbox[t]{5.6cm}{\raggedright Which passage best helps to find information about the query or to answer it?\strut} & 0.492 & $+$0.002 [$-$0.003, $+$0.006] & 0.451 & $-$0.001 [$-$0.008, $+$0.007] \\
\parbox[t]{5.6cm}{\raggedright Which passage best answers the query?\strut} & 0.491 & $+$0.001 [$-$0.006, $+$0.007] & 0.447 & $-$0.006 [$-$0.013, $+$0.002] \\
\parbox[t]{5.6cm}{\raggedright Which passage would a domain expert pick as the most useful for answering the query?\strut} & 0.500 & $+$0.009 [$+$0.003, $+$0.016] & 0.457 & $+$0.005 [$-$0.004, $+$0.014] \\
\parbox[t]{5.6cm}{\raggedright Which passage contains the knowledge needed to answer or solve the query, even if it uses different words than the query?\strut} & 0.492 & $+$0.001 [$-$0.006, $+$0.009] & 0.460 & $+$0.008 [$-$0.002, $+$0.018] \\
\parbox[t]{5.6cm}{\raggedright Which passage is the most relevant to the query, meaning it contains information that answers, supports or contradicts it?\strut} & 0.495 & $+$0.005 [$+$0.001, $+$0.009] & 0.456 & $+$0.004 [$-$0.004, $+$0.012] \\
\bottomrule
\end{tabular}}
\caption{Prompt test for the multiple-choice question on seven sets. The rows are in a fixed order, starting with the default prompt that is used everywhere else in the paper. The best prompt was chosen on a fixed random half of every set's queries. That was the fourth row (domain expert) for Perplexity v1.2 and the fifth row (different words) for Perplexity Flash v1.2. All prompts are reported here on the other half, with the mean \ndcg{} over the seven sets plus the difference to the default with its 95\% interval.}
\label{tab:prompt}
\end{table}

\section{Served model versions}
\label{app:versions}

Table~\ref{tab:versions} lists the model versions that the APIs reported on 3 November 2026, or on 13 November 2026 for Perplexity v1.3 as well as Perplexity Flash v1.3. The self-hosted models have no served version, and the other API models, which joined the main runs after 3 November, were not checked. Google's semantic ranker was called under the name in the last row, which pointed to version 005 at the time of the runs~\cite{google2026ranking}.

\begin{table}[h]
\centering
\footnotesize
\begin{tabular}{ll}
\toprule
Model & Served model \\
\midrule
Perplexity v1.2 & \texttt{pplx-decider-v1.2-20261016} \\
Perplexity Flash v1.2 & \texttt{pplx-decider-flash-v1.2-20261016} \\
Perplexity v1.3 & \texttt{pplx-decider-v1.3-20261112} \\
Perplexity Flash v1.3 & \texttt{pplx-decider-flash-v1.3-20261112} \\
GPT-6 Luna & \texttt{gpt-6-luna-decisions-20261006} \\
Jev 1.14 & \texttt{jev-1.14-20261014} \\
Liquid d1-ultra & \texttt{d1-ultra-20261015} \\
Liquid d1.1 & \texttt{d1.1-20261027} \\
joey-1 & \texttt{joey-1-20261029} \\
elephant-1 & \texttt{elephant-1-20261021} \\
Cohere Rerank 4 Pro & \texttt{rerank-v4.0-pro} \\
Voyage Rerank 2.5 & \texttt{rerank-2.5} \\
Qwen3 Reranker 8B & \texttt{qwen3-reranker-8b} (Fireworks) \\
Nemotron 3.5 Rerank 8B & \texttt{nemotron-3.5-rerank-8b} (NVIDIA) \\
Google's semantic ranker & \texttt{semantic-ranker-default@latest} (version 005) \\
\bottomrule
\end{tabular}
\caption{Served model versions.}
\label{tab:versions}
\end{table}

\section{Data}
\label{app:data}

The files published with this paper contain the per-set numbers from which every table and figure is computed, namely \ndcg{}, cost as well as latency for each system on each set, plus the bootstrap samples of the paired tests. The benchmark data are available from their original sources, and Section~\ref{sec:setup} describes how the candidate lists were built.

\end{document}